\documentclass[a4paper,12pt]{article}

\usepackage{amsmath}
\usepackage{amssymb}
\usepackage[dvips]{graphicx}
\usepackage[dvips]{psfrag}
\usepackage{bm}

\newcommand{\id}{{\bf 1}} 
\newcommand {\be}{\begin{equation}}
\newcommand {\ee}{\end{equation}}
\newcommand {\bea}{\begin{eqnarray}}
\newcommand {\eea}{\end{eqnarray}}
\newcommand {\nn}{\nonumber}
\newcommand {\tr}{{\rm tr}}
\newcommand {\Tr}{\mbox{Tr}}

\newcommand {\dd}{\mbox{d}}
\newcommand {\der}{\partial}

\newcommand{\cN}{{\cal N}}
\newcommand{\cO}{{\cal O}}

\newcommand{\ket}{\rangle}
\newcommand{\bra}{\langle}
\newcommand{\vev}[1]{\left\langle #1 \right\rangle}

\newcommand{\wt}{\widetilde}

\begin{document}
\thispagestyle{empty} \addtocounter{page}{-1}
\begin{flushright}
OIQP-09-10\\
%
\end{flushright} 
\vspace*{1cm}

\begin{center}
{\large \bf  Spontaneous supersymmetry breaking in large-$N$ matrix models \\
with slowly varying potential}\\
\vspace*{2cm}
Tsunehide Kuroki$^*$ and Fumihiko Sugino$^\dagger$\\
\vskip0.7cm
{}$^*${\it Department of Physics, Rikkyo University, }\\
\vspace*{1mm}
{\it Nishi-Ikebukuro, Tokyo 171-8501, Japan}\\
\vspace*{0.2cm}
{\tt tkuroki@rikkyo.ac.jp}\\
\vskip0.4cm
{}$^\dagger${\it Okayama Institute for Quantum Physics, } \\
\vspace*{1mm}
{\it Kyoyama 1-9-1, Kita-ku, Okayama 700-0015, Japan}\\
\vspace*{0.2cm}
{\tt fumihiko\_sugino@pref.okayama.lg.jp}\\
\end{center}
\vskip2cm
\centerline{\bf Abstract}
\vspace*{0.3cm}
{\small 
We construct a class of matrix models, where supersymmetry (SUSY) is spontaneously broken at the  
matrix size $N$ infinite. 
The models are obtained by dimensional reduction of matrix-valued SUSY quantum mechanics. 
The potential of the models is slowly varying, and the large-$N$ limit 
is taken with the slowly varying limit. 

First, we explain our formalism, introducing an external field to detect spontaneous SUSY breaking, 
analogously to ordinary (bosonic) symmetry breaking. 
It is observed that SUSY is possibly broken even in systems in less than one-dimension, 
for example, discretized quantum mechanics with a finite number of discretized time steps. 
Then, we consider spontaneous SUSY breaking in the SUSY matrix models with slowly varying potential, 
where the external field is turned off after the large-$N$ and slowly varying limit, 
analogously to the thermodynamic limit 
in statistical systems. 
On the other hand, without taking the slowly varying limit, 
in the SUSY matrix model with a double-well potential 
whose SUSY is broken due to instantons for finite $N$, 
a number of supersymmetric behavior is explicitly seen at large $N$. 
It convinces us that the instanton effect disappears and the SUSY gets 
restored in the large-$N$ limit.

}
\vspace*{1.1cm}



\newpage

\section{Introduction}
Motivated by possibility of supersymmetry (SUSY) breaking in large-$N$ matrix-model formulations of superstring 
theory~\cite{Banks:1996vh,Ishibashi:1996xs,Dijkgraaf:1997vv} (in particular \cite{Ishibashi:1996xs}), 
we presented some concrete models, where SUSY is preserved 
in any finite size $N$ of the matrix variables, 
but gets spontaneously broken at infinite $N$~\cite{Kuroki:2007iy}. 
They are in the form of matrix models coupled to some supersymmetric field theories. 
The supersymmetric field theories alone undergo spontaneous SUSY breaking, however the couplings to the 
matrix model sector supply some effect to prevent the breaking. 
Thus, the models for finite $N$ preserve the SUSY. 
As $N$ gets large, the effect diminishes, 
and eventually nothing prevents the SUSY breaking in the field theory 
sector at $N$ infinite. In this mechanism, the models exhibit the SUSY breaking only at the large-$N$ limit. 

{}From our motivation, it is certainly desirable to construct models possessing the above property 
within the framework of a matrix model, 
not relying on SUSY breaking in the field theory sector. 
In this paper, we construct a class of such desirable large-$N$ SUSY matrix models.
They are obtained by dimensional 
reduction of matrix-valued supersymmetric quantum mechanics with $\cN=2$ SUSY. 
It is expected that they play some role to reveal the essence of SUSY breaking of nonperturbative formulation 
of string theory via matrix models.

In the next section, analogously to the situation of ordinary spontaneous symmetry breaking, 
we introduce an external field to choose one of degenerate broken vacua 
to detect spontaneous SUSY breaking in SUSY quantum mechanics. 
The external field plays the same role as  
the magnetic field in the Ising model introduced to detect the magnetization. 
To introduce a suitable external field for the supersymmetric system, 
we deform the boundary condition for fermions from the periodic boundary condition 
(PBC) to a twisted boundary condition (TBC) with twist $\alpha$,  
which can be regarded as the external field. 
If a supersymmetric system undergoes spontaneous SUSY breaking, 
the partition function with the PBC for all the fields, 
$Z_{\rm PBC}$, which usually corresponds to the Witten index~\cite{Witten:1982df},  
is expected to vanish. 
Then, the expectation values of observables, which are normalized by $Z_{\rm PBC}$, 
would be ill-defined or indefinite.  
By introducing the twist, the partition function is regularized and the expectation values become 
well-defined. 
It is an interesting aspect of our external field for SUSY breaking, 
which is not seen in spontaneous breaking of ordinary (bosonic) symmetry.  

Differently from quantum field theory whose degrees of freedom are infinite, 
one-dimensional SUSY quantum mechanics has finite degrees of freedom, 
and thus the superselection rule does not hold in general. 
Here, introducing the twist $\alpha$ enables us to show that, 
concerning the expectation values of the auxiliary fields, 
which are order 
parameters of the SUSY breaking, the superselection rule works and the expectation values at the one broken 
ground state are well-defined.   
On the other hand, we see that the expectation value of the scalar field becomes ill-defined 
as the external field $\alpha$ is turned off.   

Furthermore, it is straightforward to apply our argument to systems in less than one-dimension, 
for example   
discretized SUSY quantum mechanics with a finite number of discretized time steps. 
Spontaneous SUSY breaking is observed even in such simple systems with lower degrees of freedom. 
Also, we give some argument that an analog of 
the Mermin-Wagner-Coleman theorem~\cite{Mermin:1966fe,Coleman:1973ci} 
does not hold for SUSY. 
Thus, cooperative phenomena are not essential to cause spontaneous SUSY breaking, 
which makes a difference from spontaneous breaking of the ordinary (bosonic) symmetry~\footnote{
It has already been pointed out in ref.~\cite{Witten:1981nf}. 
Concerning this issue, 
a new ingredient in this paper is SUSY breaking in systems in less than one-dimension.}.       
As for the superselection rule, the situation is similar to the SUSY quantum mechanics. 
However, for large-$N$ SUSY matrix models discussed from section~\ref{sec:MM}, 
we can see that the superselection rule does hold in general thanks to the large-$N$ limit.  

In section~\ref{sec:MM}, we define a matrix model analog of the discretized SUSY quantum mechanics 
with general superpotential, 
and give an overview of its general aspects in the large-$N$ limit. 
We mainly discuss the simplest case with the number of the time step $T=1$. 
To realize SUSY breaking at large $N$, 
we put a parameter $\epsilon$ into the superpotential. 
$1/\epsilon$ represents the length scale of the scalar $\phi$ in which  
the derivative of the superpotential $W'(\phi)$ varies by the amount of $\cO(1)$. 
Thus, as $\epsilon$ gets smaller, the potential becomes 
more slowly varying.  
If we take the large-$N$ limit with the slowly varying limit $\epsilon \to 0$,  
the system is shown to exhibit spontaneous SUSY breaking. 
$\alpha$ is an infinitesimal external field which slightly breaks the SUSY, 
and we observe whether its breaking effect remains 
after the limit $\alpha\to 0$ following the large-$N$ and slowly varying limit, 
which is analogous to detect spontaneous symmetry breaking in the thermodynamic limit 
of a statistical system. 
As a deformation of the model, when the twist $\alpha$ is changed to a field-dependent function, 
the SUSY is also spontaneously broken in the same limit. 
The eigenvalue distribution of $\phi$ does not qualitatively change by the deformation, 
but its precise form is modified. 
Since the deformation effect remains even after turning off the external field, 
we see that the system in the large-$N$ limit with $\epsilon\to 0$ is somewhat sensitive to the 
deformation of the external field. 
Thus, as a complete definition of the model, we should specify not only the matrix-model action 
but also the form of the external field.   
This is reasonable because the twist modifies the boundary condition of the fermions, 
which is a part of the definition of the model even if the twist is turned off eventually.

In section~\ref{sec:phi4_MM}, we discuss one example of the models discussed in section~\ref{sec:MM},  
where the scalar potential is in a double-well type ($W'(\phi)$ is quadratic). 
Ref.~\cite{Witten:1982df} discusses the two-dimensional Wess-Zumino model 
with such a double-well potential. 
For finite spatial volume, the instanton effect causes SUSY breaking, while the effect ceases in the 
infinite volume limit to restore the SUSY. 
Analogously, in our matrix model case, 
SUSY will be spontaneously broken due to some sort of instanton effect for finite $N$. 
However, if we take the large-$N$ limit with $\epsilon$ kept finite (not slowly varying), 
the instanton effect is expected to disappear 
and the SUSY will be restored. 
By examining large-$N$ solutions of the matrix model, we observe a number of supersymmetric behavior 
that strongly indicates the restoration of the SUSY 
at the large-$N$ limit. 
To best of our knowledge, this is the first example of a matrix model with the restoration of SUSY 
at large $N$, although Ref.~\cite{Affleck:1983pn} discusses it for a large-$N$ vector model. 

In the SUSY matrix model with a double-well potential, we find an intriguing fact that 
the free energy does not depend on the ratio of the numbers of the eigenvalues 
filled into the two wells, 
and the value of the free energy is equal to that of the Gaussian SUSY matrix model. 
We think that it is deeply connected to the supersymmetric properties of the model, and 
further analysis will be reported in the next publication. 

On the other hand, we also consider the large-$N$ and slowly varying limit for the same model. 
Then, solutions which 
exhibit spontaneous SUSY breaking are obtained. 
 
In section~\ref{sec:phi6_MM}, we consider the SUSY matrix model with a cubic $W'(\phi)$,  
where SUSY is expected to be preserved for finite $N$. 
It is found that spontaneous SUSY breaking takes place in the large-$N$ limit with $\epsilon\to 0$ 
even in this model. 

In section~\ref{sec:gaussian_MM}, we discuss on the SUSY matrix model with slowly varying 
Gaussian potential. Although the simple twist does not lead to SUSY breaking, 
deformations to $\phi$-dependent twists cause SUSY breaking in the large-$N$ and slowly varying limit. 

Section~\ref{sec:summary} is devoted to summarize the results obtained so far and 
to discuss some future subjects.  
 
The path-integral expression of the partition function for SUSY quantum mechanics 
with twist $\alpha$ is derived in 
appendix~\ref{app:TPF}, and appendix~\ref{app:gaussian} is devoted to some fundamental computation 
for the Gaussian SUSY matrix model. Finally, we give a proof of key equations 
concerning the SUSY restoration of 
the double-well matrix model at large-$N$ in appendix~\ref{app:proof}.

\section{Preliminaries on SUSY quantum mechanics} 
\label{sec:SQM}
\setcounter{equation}{0}
As a preparation to discuss large-$N$ SUSY matrix models, 
in this section we present some preliminary results on SUSY quantum mechanics. 

Let us start with a system defined by the Euclidean (Wick-rotated) action: 
\be
S=\int_0^\beta dt\,  \left[\frac12 B^2 +iB\left(\dot{\phi} + W'(\phi)\right) 
+\bar{\psi}\left(\dot{\psi} +W''(\phi)\psi\right)\right], 
\label{S_SUSYQM}
\ee
where $\phi$ is a real scalar field, $\psi, \bar{\psi}$ are fermions, and $B$ is an auxiliary field. 
The dot means the derivative with respect to the Euclidean time $t \in [0, \beta]$.  
For a while, all the fields are supposed to obey the PBC. 
$W(\phi)$ is a real function of $\phi$ called superpotential, and the prime ($'$) represents 
the $\phi$-derivative. 

$S$ is invariant under one-dimensional $\cN=2$ SUSY transformations generated 
by $Q$ and $\bar{Q}$. They act on the fields as  
\bea
Q\phi =\psi, & & Q\psi=0, \nn \\
Q\bar{\psi} =-iB, & & QB=0, 
\label{QSUSY}
\eea
and 
\bea
\bar{Q} \phi = -\bar{\psi}, & & \bar{Q}\bar{\psi} = 0, \nn \\
\bar{Q} \psi = -iB +2\dot{\phi}, & & \bar{Q} B = 2i\dot{\bar{\psi}}, 
\label{Qbar_SUSY}
\eea
with satisfying the algebra
\be
Q^2=\bar{Q}^2=0, \qquad \{ Q, \bar{Q}\} =2\der_t . 
\label{SUSY_alg}
\ee
Note that $S$ can be written as the $Q$- or $Q\bar{Q}$-exact form: 
\bea
S & = &  Q\int dt \, \bar{\psi}\left\{ \frac{i}{2} B -\left(\dot{\phi} +W'(\phi)\right)\right\}  \\
  & = & Q\bar{Q} \int dt \, \left( \frac12\bar{\psi}\psi +W(\phi)\right). 
\label{S_SUSYQM2}
\eea

For demonstration, let us consider the case of the derivative of the superpotential 
\be
W'(\phi) = g(\phi^2 +\mu^2).
\label{W}
\ee
For $\mu^2 >0$, the classical minimum is given by the static configuration $\phi=0$, with its energy 
$E_0 =\frac12 g^2 \mu^4 >0$ implying spontaneous SUSY breaking. 
Then, $B=-ig\mu^2\neq 0$ from the equation of motion, 
leading to $Q\bar{\psi}, \bar{Q}\psi \neq 0$, which also means 
the SUSY breaking. 

For $\mu^2 <0$, the classical minima $\phi = \pm\sqrt{-\mu^2}$ are zero-energy configurations. 
It is known that the quantum tunneling (instantons) between the minima resolves 
the degeneracy giving positive energy to the ground state.  
SUSY is broken also in this case. 

Next, let us consider quantum aspects of the SUSY breaking in this model. 
For later discussions on matrix models, it is desirable to observe SUSY breaking 
via the path integral formalism, that is, by seeing the expectation value of some field. 
We take $\vev{B}$ (or $\vev{B^n}$ ($n=1,2,\cdots$)) as such an order parameter. 
Whichever $\mu^2$ is positive or negative, the SUSY is broken, so 
the ground state energy $E_0$ is positive. Then, 
for each of the energy levels $E_n$ ($0<E_0<E_1<E_2<\cdots$), 
the SUSY algebra\footnote{In the operator formalism, 
$\bar{Q}$, $\bar{\psi}$ are regarded as hermitian conjugate to $Q$, $\psi$, respectively.}   
\be
\{ Q, \bar{Q} \} = 2E_n, \qquad Q^2 = \bar{Q}^2 =0 
\ee
leads to the SUSY multiplet formed by bosonic and fermionic states
\be
|b_n\ket = \frac{1}{\sqrt{2E_n}}\,\bar{Q}|f_n\ket, \qquad 
|f_n\ket = \frac{1}{\sqrt{2E_n}} \,Q |b_n\ket.
\ee
As a convention, we assume that $|b_n\ket$ and $|f_n\ket$ 
have the fermion number charges $F= 0$ and $1$, respectively. 
Since the $Q$-transformation for $B$ in (\ref{QSUSY}) is expressed as 
\be
[Q, B] =0 
\ee
in the operator formalism, we can see that 
\be
\bra b_n |B|b_n\ket = \bra f_n|B|f_n\ket 
\label{BVEV}
\ee
holds for each $n$. 
Then, it turns out that the unnormalized expectation value of $B$ vanishes\footnote{Furthermore, 
$\bra B^n\ket' =0$ $(n=1,2,\cdots)$ can be shown.}: 
\bea
\bra B\ket'  & \equiv & \int _{\rm PBC} \dd (\mbox{fields}) \, B \, e^{-S}  \nn \\
                & = & \Tr \left[B (-1)^F e^{-\beta H}\right]  \nn \\
                & = & \sum_{n=0}^\infty\left(\bra b_n |B|b_n\ket - \bra f_n|B|f_n\ket \right) e^{-\beta E_n} 
\nn \\
 & = & 0. 
\label{naiveBVEV}                
\eea
This observation shows that, in order to judge SUSY breaking from the expectation value of $B$, 
we should choose either of the SUSY broken ground states ($|b_0\ket$ or $|f_0\ket$) and 
see the expectation value with respect to the chosen ground state. 
It makes sense, because there is no transition between the two ground states via an arbitrary bosonic 
operator $\cO_B$ (including $B$) from the conservation of the $(-1)^F$-charge: 
\be
\bra b_0 |\cO_B|f_0\ket =0. 
\label{selection}
\ee
The situation is somewhat analogous to the case of spontaneous breaking of ordinary (bosonic) symmetry 
in the sense that \eqref{selection} is reminiscent of the superselection rule.

However, differently from the ordinary case, when SUSY is broken, the supersymmetric 
partition function vanishes:
\bea
Z_{\rm PBC} & = & \int _{\rm PBC} \dd (\mbox{fields})  \, e^{-S}   =  \Tr \left[(-1)^F e^{-\beta H}\right] 
\label{partition_fun_PBC}  \\
 & = & \sum_{n=0}^\infty \left(\bra b_n |b_n\ket - \bra f_n|f_n\ket \right) e^{-\beta E_n} \nn \\
 & = & 0,
\eea
where the normalization $\bra b_n |b_n\ket = \bra f_n|f_n\ket =1$ was used. 
So, the expectation values normalized by $Z_{\rm PBC}$ could be ill-defined~\cite{kanamori-ss}.

\subsection{Twisted boundary condition}
To detect spontaneous breaking of ordinary symmetry, 
some external field is introduced so that  
the ground state degeneracy is resolved to specify a single broken ground state. 
The external field is turned off after 
taking the thermodynamic limit, then we can judge whether spontaneous 
symmetry breaking takes place or not, seeing the value of the corresponding order parameter.   
(For example, to detect the spontaneous magnetization in the Ising model, 
the external field is a magnetic field, and the corresponding order 
parameter is the expectation value of the spin operator.)   

We will do a similar thing also for the case of spontaneous SUSY breaking. 
For this purpose, let us change the boundary condition for the fermions to the TBC: 
\be
\psi(t+\beta) = e^{i\alpha} \psi(t), \qquad 
\bar{\psi}(t+\beta) = e^{-i\alpha}\bar{\psi}(t), 
\label{twistedBC}
\ee
then the twist $\alpha$ can be regarded as an external field. 
Other fields remain intact. 
As seen in appendix~A, the partition function with the TBC corresponds to 
the expression (\ref{partition_fun_PBC}) with $(-1)^F$ replaced by $(-e^{-i\alpha})^F$: 
\bea  
Z_\alpha & \equiv & -e^{-i\alpha}\int _{\rm TBC} \dd (\mbox{fields})  \, e^{-S}  
 =  \Tr \left[(-e^{-i\alpha})^F e^{-\beta H}\right]  
\label{partition_fun_TBC} \\
 & = & \sum_{n=0}^\infty \left(\bra b_n |b_n\ket - e^{-i\alpha}\bra f_n|f_n\ket \right) e^{-\beta E_n} \nn \\
 & = & \left(1-e^{-i\alpha}\right) \sum_{n=0}^\infty e^{-\beta E_n}. 
\eea

Then, the normalized expectation value of $B$ under the TBC becomes 
\bea
\vev{B}_\alpha &\equiv & \frac{1}{Z_\alpha} \,\Tr\left[B (-e^{-i\alpha})^F e^{-\beta H}\right] \nn \\
    & = & \frac{1}{Z_\alpha}\, \sum_{n=0}^\infty \left(\bra b_n| B |b_n\ket - e^{-i\alpha} \bra f_n| B |f_n\ket
\right) e^{-\beta E_n}\nn \\
 & = & \frac{\sum_{n=0}^\infty \bra b_n| B |b_n\ket e^{-\beta E_n}}{\sum_{n=0}^\infty e^{-\beta E_n}}
= \frac{\sum_{n=0}^\infty \bra f_n| B |f_n\ket e^{-\beta E_n}}{\sum_{n=0}^\infty e^{-\beta E_n}}.
\label{vev_B_TBC} 
\eea
Note that the factors $\left(1-e^{-i\alpha}\right)$ in the numerator and the denominator  
cancel each other, and thus $\vev{B}_\alpha$ does not depend on $\alpha$ even for finite $\beta$. 
As a result, $\vev{B}_\alpha$ is equivalent to the expectation value taken over 
one of the ground states and its excitations $\{ |b_n\ket \}$ (or $\{|f_n\ket \}$). 
The normalized expectation value of $B$ under the PBC was of the indefinite form $0/0$, 
which is now regularized by introducing the parameter $\alpha$. 
The expression (\ref{vev_B_TBC}) is well-defined. 

On the other hand, from the $Q$-transformation $\psi = [Q, \phi]$, we have 
\be
\bra b_n|\phi|b_n\ket = \bra f_n|\phi|f_n\ket + \frac{1}{\sqrt{2E_n}} \bra f_n|\psi|b_n\ket. 
\label{transition_b_f}
\ee
The second term is a transition between bosonic and fermionic states via the fermionic operator $\psi$, 
which does not vanish in general. 
Thus, differently from $\vev{B}_\alpha$, the expectation value of $\phi$ becomes 
\bea
\vev{\phi}_\alpha & = & \frac{1}{Z_\alpha} \,\Tr\left[\phi (-e^{-i\alpha})^F e^{-\beta H}\right] \nn \\
   & = & \frac{1}{Z_\alpha} \, \sum_{n=0}^\infty 
\left(\bra b_n|\phi|b_n\ket -e^{-i\alpha}\bra f_n|\phi|f_n\ket\right) e^{-\beta E_n} \nn \\
 & = & \frac{\sum_{n=0}^\infty \bra f_n| \phi |f_n\ket e^{-\beta E_n}}{\sum_{n=0}^\infty e^{-\beta E_n}} 
 + \frac{1}{1-e^{-i\alpha}}\,\frac{\sum_{n=0}^\infty \bra f_n| \psi |b_n\ket \frac{1}{\sqrt{2E_n}} e^{-\beta E_n}}{\sum_{n=0}^\infty e^{-\beta E_n}}. 
 \label{phi_vev_alpha_QM}
\eea 
When $\bra f_n| \psi |b_n\ket\neq 0$ for some $n$, 
the second term is $\alpha$-dependent and diverges as $\alpha\to 0$. 
The divergence comes from the transition between $|b_n\ket$ and $|f_n\ket$. 
Since the two states are transformed to each other by the (broken) SUSY transformation, 
we can say that they should belong to the separate superselection sectors, 
in analogy to spontaneous breaking of ordinary (bosonic) symmetry. 
Thus, the divergence of $\vev{\phi}_\alpha$ as $\alpha\to 0$ implies that 
the superselection rule does not hold in the system~\footnote{
For higher dimensional systems defined on finite spatial volume, 
such phenomena would also happen in general.}.

\subsection{Discretized SUSY quantum mechanics}
\label{sec:discrete_phi4}
In this subsection, we consider a discretized system of (\ref{S_SUSYQM}), namely the Euclidean time 
is discretized as $t=1,\cdots,T$. 
The action is written as 
\bea
S & = & Q \sum_{t=1}^T \bar{\psi}(t) \left\{ \frac{i}{2} B(t) 
-\left(\phi(t+1) -\phi(t) +W'(\phi(t))\right)\right\} 
\label{dQM_S}\\
 & = & \sum_{t=1}^T \left[\frac12 B(t)^2 +iB(t)\left\{\phi(t+1)-\phi(t)+W'(\phi(t))\right\} \right. \nn \\
 & & \hspace{2cm} \left. \frac{}{} +\bar{\psi}(t) \left\{\psi(t+1) -\psi(t) +W''(\phi(t)) \psi(t)\right\} \right],  
 \label{dQM_S2}
\eea
where the $Q$-supersymmetry is of the same form as in (\ref{QSUSY}). 
As is seen by the $Q$-exact form (\ref{dQM_S}), the action is $Q$-invariant and the $Q$-supersymmetry is preserved 
upon the discretization~\cite{catterall4}. 
On the other hand, the $\bar{Q}$-supersymmetry can not be preserved by the discretization in the case of  
$T\geq 2$. 

When $T$ is finite, the partition function or various correlators are expressed as   
a finite number of integrals with respect to field variables. 
So, at first sight, one might expect that spontaneous breaking of the 
SUSY could not take place, because of a small number of the degrees of freedom. 
In what follows, we will demonstrate that the expectation is not correct, 
and that the SUSY can be broken even in such a finite system.  

\subsubsection{$T=1$ case}
\label{subsub:T=1}
First, let us consider the simplest case\footnote{For $T=1$ with the PBC, the action is same as the 
dimensional reduction of (\ref{S_SUSYQM}). Then, the $\bar{Q}$-supersymmetry 
\bea
\bar{Q} \phi = -\bar{\psi}, & & \bar{Q}\bar{\psi} = 0, \nn \\
\bar{Q} \psi = -iB, & & \bar{Q} B = 0 
\label{Q_bar_T=1}
\eea
becomes the symmetry of the action again.} $T=1$.  
As before, the boundary condition of the fermion is twisted with the phase $\alpha$:  
\be
\phi(2) = \phi(1), \qquad \psi(2) = e^{i\alpha} \psi(1). 
\label{TBC_alpha}
\ee
The partition function 
\bea
Z_\alpha & \equiv & \frac{-1}{2\pi}\int d B \,d \phi \,d \psi \,d \bar{\psi} \, e^{-S_\alpha}, \\
S_\alpha & = & \frac12 B^2 +iB W'(\phi) +\bar{\psi} \left(e^{i\alpha}-1 +W''(\phi)\right) \psi
\eea
with the superpotential (\ref{W}) is computed to be 
\bea
Z_\alpha & = & \frac{1}{\sqrt{2\pi}} \int^\infty_{-\infty}  d\phi \, \left(e^{i\alpha}-1 +W''(\phi)\right) e^{-\frac12 W'(\phi)^2} \nn \\
            & = &  \left(e^{i\alpha}-1 \right)  C, \\
         C & \equiv  &    \frac{1}{\sqrt{2\pi}}\int^\infty_{-\infty}  d\phi \, e^{-\frac12 W'(\phi)^2} . 
\eea
$C$ is positive definite, 
and it is same as in the continuum case that $Z_\alpha$ approaches to zero as $\alpha \to 0$. 

An analog to the unnormalized expectation value (\ref{naiveBVEV}) with the PBC is 
\bea
\vev{B}' & \equiv & \frac{-1}{2\pi} \int d B \,d \phi \,d \psi \,d \bar{\psi} \, B \, e^{-S_{\alpha=0}} \nn \\
           & = & \frac{-i}{\sqrt{2\pi}}\int d \phi \, W'(\phi) W''(\phi) \, e^{-\frac12 W'(\phi)^2} \nn \\
           & = & \frac{i}{\sqrt{2\pi}}\int^\infty_{-\infty}  d\phi \, \frac{\der}{\der\phi} \, e^{-\frac12 W'(\phi)^2} =0. 
\eea
However, the normalized expectation value $\vev{B}_\alpha$ becomes  
\bea
\vev{B}_\alpha & = & \frac{1}{Z_\alpha}\frac{-1}{2\pi} \int d B \,d \phi \,d \psi \,d \bar{\psi} \, B \, e^{-S_\alpha}   \nn \\
   & = & \frac{1}{Z_\alpha}\frac{-i}{\sqrt{2\pi}} \int^\infty_{-\infty}  d\phi \, W'(\phi) \left(e^{i\alpha}-1 +W''(\phi)\right) 
                e^{-\frac12 W'(\phi)^2}  \nn \\
   & = & \frac{1}{Z_\alpha} \left(e^{i\alpha}-1\right) \frac{-i}{\sqrt{2\pi}}   \int^\infty_{-\infty} d\phi \,
                  W'(\phi)  e^{-\frac12 W'(\phi)^2}    \nn \\
   & = & \frac{-i}{\sqrt{2\pi}} \frac{1}{C}\int^\infty_{-\infty} d\phi \,W'(\phi)  e^{-\frac12 W'(\phi)^2} . 
\label{BVEV_alpha}   
\eea
Again, 
the factors $\left(e^{i\alpha}-1\right)$ from the numerator and the denominator cancel each 
other, and the result does not depend on $\alpha$. 
For $\mu^2>0$, (\ref{BVEV_alpha}) is not zero, implying SUSY breaking. 
At the classical level, $B=-ig\mu^2$ from the equation of motion, 
which coincides to (\ref{BVEV_alpha}) evaluated at the potential minimum $\phi=0$. 
This shows that the introduction of the external field $\alpha$ gives a correct regularization. 
Also, for $\mu^2<0$, generically (\ref{BVEV_alpha}) does not vanish\footnote{Because 
\be
B^n = Q\left(i\bar{\psi}B^{n-1}\right) \qquad (n=1,2,\cdots), 
\ee 
we can conclude spontaneous SUSY breaking by showing $\vev{B^n}_\alpha\neq 0$ 
for some $n$. 
Even when $\vev{B}_\alpha =0$ holds accidentally for $\mu^2<0$, 
we could show $\vev{B^n}_\alpha\neq 0$ for some $n$ and thus SUSY breaking.}, 
and it is found that the SUSY is spontaneously broken. 

The unnormalized expectation value with the TBC  
\be
\vev{B}'_\alpha = 
\left(e^{i\alpha}-1\right)  \frac{-i}{\sqrt{2\pi}} \int^\infty_{-\infty} d\phi \,W'(\phi) 
e^{-\frac12 W'(\phi)^2}  
\label{Bvev'_TBC}
\ee
vanishes in the limit $\alpha\to 0$, and no singular behavior can be seen there. 
For the normalized expectation value, however, both of the numerator $\vev{B}'_\alpha$ and 
the denominator $Z_\alpha$ approach to zero as $\alpha \to 0$. 
The dependence of $\alpha$ is canceled between the numerator and the denominator, and then 
the nonvanishing result arises. 
In other words, the external field $\alpha$ regularizes the indefinite form of the expectation value 
under the PBC: $\vev{B}=0/0$, and leads to the nontrivial result. 
Thus, we conclude that, even in discrete systems defined by a finite number of integrals, 
SUSY can be spontaneously broken when the partition function vanishes. 
Note that it is totally different from the spontaneous breaking of ordinary (bosonic) symmetry, 
where the partition function never vanishes. 
In general, some non-analytic behavior is necessary for spontaneous symmetry breaking. 
For the SUSY breaking in the finite system, 
the non-analyticity is supplied by the vanishing partition function.    

As for the expectation value $\vev{\phi}_\alpha$, similarly to the case 
(\ref{phi_vev_alpha_QM}), we obtain
\bea
\vev{\phi}_\alpha & = & \frac{1}{Z_\alpha}\frac{-1}{2\pi}
\int d B \,d \phi \,d \psi \,d \bar{\psi} \, \phi \, e^{-S_\alpha}   \nn \\
   & = & \frac{1}{e^{i\alpha}-1}\frac{1}{\sqrt{2\pi}} \frac{1}{C} \int^\infty_{-\infty}  d\phi \, \phi W''(\phi)
                e^{-\frac12 W'(\phi)^2} . 
\label{phi_vev_alpha_dQM}                
\eea
Since $\phi W''(\phi) = 2g\phi^2$, the integral of the right-hand side does not vanish, and $\vev{\phi}_\alpha$ 
diverges as $\alpha\to 0$. 

In what follows from section~\ref{sec:MM}, 
we will consider various large-$N$ matrix models analogous to the models presented 
so far. Interestingly, the singular behavior of $\vev{\phi}_\alpha$ as seen above does not appear there. 
It can be understood that the tunneling between separate broken vacua is suppressed 
by taking the large-$N$ limit, and thus the superselection rule works.  
Note that the large-$N$ limit in the matrix models is analogous to the infinite volume limit or 
the thermodynamic limit of statistical systems.     
In fact, this will play an essential role 
for restoration of SUSY in the large-$N$ limit of the matrix model with a double-well potential.     

\paragraph{General $W'(\phi)$ case}                                
Similar analyses can be done for $W'(\phi)$ given as a general polynomial of 
the degree $k$: 
\be
W'(\phi)= g_k \phi^k + g_{k-1} \phi^{k-1} + \cdots + g_0.
\label{poly}
\ee 
Because 
\be
\frac{1}{\sqrt{2\pi}}\int^\infty_{-\infty} d\phi \, W''(\phi) \, e^{-\frac12 W'(\phi)^2} 
= \left\{ \begin{array}{cl} {\rm sgn}(g_k) & \mbox{for $k$: odd} \\
                                 0 & \mbox{for $k$: even} \end{array}\right.
\label{dQM_generalW'}                                 
\ee
(from the Nicolai mapping~\cite{Nicolai:1979nr}), and (\ref{Bvev'_TBC}) holds 
for general $W'(\phi)$, we see that the result of $\vev{B}_\alpha$ does not change from (\ref{BVEV_alpha}) 
for even $k$, while 
$\lim_{\alpha \to 0}\vev{B}_\alpha=0$ for odd $k$. 
(Note that $\lim_{\alpha\rightarrow 0}Z_\alpha\neq 0$
for odd $k$, 
since (\ref{dQM_generalW'}) is nonzero.) 
Thus, for the discretized system with the superpotential $W'(\phi)$ 
of the degree even (odd), the SUSY is broken (preserved), which is same as the conclusion for the 
continuous SUSY quantum mechanics~\cite{Witten:1981nf}.

\subsubsection{General $T$ case}
It is straightforward to extend the above discussion to the case of general $T$. 

Expressing as $S_\alpha$ the action (\ref{dQM_S2}) under the TBC 
\be
\phi(T+1) = \phi(1), \qquad \psi(T+1) = e^{i\alpha} \psi(1), 
\ee
the partition function 
\be
Z_\alpha \equiv \left(\frac{-1}{2\pi}\right)^T \int
\prod_{t=1}^T\left(dB(t)\,d\phi(t)\,d\psi(t)\,d\bar{\psi}(t)\right) e^{-S_\alpha}
\ee
is computed to be 
\bea
Z_\alpha  & = & (-1)^T\left(1-e^{i\alpha}\right) C_T, \\
C_T & \equiv & \int 
\left(\prod_{t=1}^T\frac{d\phi(t)}{\sqrt{2\pi}}\right) 
e^{-\frac12\sum_{t=1}^T\left(\phi(t+1)-\phi(t)+W'(\phi(t))\right)^2}.  
\eea
Here we used 
\be
\int 
\left(\prod_{t=1}^T\frac{d\phi(t)}{\sqrt{2\pi}}\right) 
\left[\prod_{t=1}^T\left(-1+W''(\phi(t))\right) -(-1)^T\right] 
e^{-\frac12\sum_{t=1}^T\left(\phi(t+1)-\phi(t)+W'(\phi(t))\right)^2} =0 
\ee
for the superpotential (\ref{W}), which is derived from the Nicolai mapping~\cite{Nicolai:1979nr}. 
(Note the factor $\left[\prod_{t=1}^T\left(-1+W''(\phi(t))\right) -(-1)^T\right]$ 
is equal to the fermion determinant under the PBC.)  
Also, $C_T$ is positive definite. 

Similarly, for the normalized expectation value 
\be
\vev{B(t)}_\alpha \equiv \frac{1}{Z_\alpha} \left(\frac{-1}{2\pi}\right)^T \int
\prod_{t=1}^T\left(dB(t)\,d\phi(t)\,d\psi(t)\,d\bar{\psi}(t)\right) \, B(t) \,e^{-S_\alpha}, 
\ee
we use the Nicolai mapping to have  
\bea
\vev{B(t)}_\alpha & = &  \frac{1}{Z_\alpha}(-1)^T\left(1-e^{i\alpha}\right) 
\int 
\left(\prod_{t=1}^T\frac{d\phi(t)}{\sqrt{2\pi}}\right) \,(-i)\left(\phi(t+1)-\phi(t)+W'(\phi(t))\right) \nn \\
 & & \hspace{6cm} \times e^{-\frac12\sum_{t=1}^T\left(\phi(t+1)-\phi(t)+W'(\phi(t))\right)^2} \nn \\
 & = & \frac{1}{C_T}
\int 
\left(\prod_{t=1}^T\frac{d\phi(t)}{\sqrt{2\pi}}\right) \,(-i)\left(\phi(t+1)-\phi(t)+W'(\phi(t))\right) \nn \\
 & & \hspace{3.5cm} \times e^{-\frac12\sum_{t=1}^T\left(\phi(t+1)-\phi(t)+W'(\phi(t))\right)^2}. 
 \label{dQM_Bvev}
\eea
The factor $(-1)^T\left(1-e^{i\alpha}\right)$ was canceled, 
and $\vev{B(t)}_\alpha$ does not depend on $\alpha$, again. 
The result (\ref{dQM_Bvev}) is finite and well-defined. 
By using the Nicolai mapping, it is straightforward to generalize this result to the case of \eqref{poly}. 
We find that \eqref{dQM_Bvev} holds and it is finite and well-defined for even $k$, 
and that $\lim_{\alpha\rightarrow 0} \vev{B(t)}_\alpha=0$ for odd $k$. 

\paragraph{No analog of Mermin-Wagner-Coleman theorem for SUSY}
As claimed in the Mermin-Wagner-Coleman theorem~\cite{Mermin:1966fe,Coleman:1973ci}, 
continuous bosonic symmetry cannot be spontaneously broken at the quantum level 
in the dimensions of two or lower. 
In dimensions $D\leq 2$, although the symmetry might be broken at the classical level, 
in computing quantum corrections to a classical (nonzero) value of a corresponding order parameter, 
one encounters IR divergences from loops of a massless boson. 
It indicates that the conclusion of the symmetry breaking from the classical value is not reliable 
at the quantum level any more. 
It is a manifestation of the Mermin-Wagner-Coleman theorem.  

Here, we consider whether an analog of the Mermin-Wagner-Coleman theorem for SUSY holds or not. 
Naively, since loops of a massless fermion~\footnote{For theories in finite volume, 
the massless fermion might be regarded as ``would-be Nambu-Goldstone fermion'', 
because the Nambu-Goldstone fermion is defined for theories in the infinite volume 
and its concept is ill-defined for finite volume~\cite{Witten:1982df}.}
would be dangerous in the dimension one or lower, 
we might be tempted to expect that SUSY could not be spontaneously broken at the quantum level 
in the dimension of one or lower. 
However, this expectation is not correct.  
Because the twist $\alpha$ in our setting can also be regarded as an IR cutoff for the massless fermion, 
the finiteness of (\ref{dQM_Bvev}) shows that $\vev{B(t)}_\alpha$ is free from IR divergences and 
well-defined at the quantum level for less than one-dimension. 
(For one-dimensional case, (\ref{vev_B_TBC}) has no $\alpha$-dependence, 
thus no IR divergences.)

We can see it more explicitly in perturbative calculations. 
Let us consider the superpotential~(\ref{W}) with $\mu^2 >0$, 
where the classical configuration $\phi(t)=0$ gives $B(t) = -ig\mu^2$. 
If the theorem holds, quantum corrections should modify this classical value to zero, 
and there we should come across IR divergences owing to a massless fermion. 
Although we have obtained the finite result \eqref{dQM_Bvev}, the following perturbative analysis 
would clarify a role played by the massless fermion. 
We evaluate quantum corrections to the classical value of $B(t)$ perturbatively. 
Under the mode expansions 
\bea
\phi(t) & = &  \frac{1}{\sqrt{T}}\sum_{n=-(T-1)/2}^{(T-1)/2} \wt{\phi}_n \, e^{i2\pi nt/T}
\qquad \mbox{with} \quad \wt{\phi}_n^* = \wt{\phi}_{-n}, \nn \\
\psi(t) & = &  \frac{1}{\sqrt{T}}\sum_{n=-(T-1)/2}^{(T-1)/2} \wt{\psi}_n \, e^{i(2\pi n +\alpha)t/T}, 
\nn \\
\bar{\psi}(t) & = & \frac{1}{\sqrt{T}}\sum_{n=-(T-1)/2}^{(T-1)/2} \wt{\overline{\psi}}_n \, e^{-i(2\pi n +\alpha)t/T},
\eea
free propagators are 
\bea
\vev{\wt{\phi}_{-n}\wt{\phi}_m}_{\rm free} & = & \frac{\delta_{nm}}{4\sin^2\left(\frac{\pi n}{T}\right) +M^2}, \nn \\
\vev{\wt{\psi}_n\wt{\overline{\psi}}_m}_{\rm free} & = & \frac{\delta_{nm}}{e^{i(2\pi n +\alpha)/T}-1}
\eea
with $M^2 \equiv 2g^2\mu^2$. Here we consider the case of odd $T$ for simplicity of the mode expansion.
Note that the boson is massive while the fermion is nearly massless regulated by $\alpha$. 
Also, there are three kinds of interactions in $S_\alpha$ (after $B$ is integrated out): 
\bea
& & V_4 = \sum_{t=1}^T\frac12 g^2 \phi(t)^4, \nn \\
& & V_{3B} = \sum_{t=1}^T g\phi(t)^2\left(\phi(t+1)-\phi(t)\right), \qquad 
V_{3F} = \sum_{t=1}^T 2g \phi(t) \bar{\psi}(t)\psi(t). 
\eea

We perturbatively compute the second term of 
\be
\vev{B(t)}_\alpha = -ig\mu^2 -i\vev{g\phi(t)^2 +\phi(t+1)-\phi(t)}_\alpha 
\ee
up to the two-loop order, and directly see that the nearly massless fermion 
(``would-be Nambu-Goldstone fermion'') does not contribute and gives no IR singularity. 
It is easy to see that the tadpole $\vev{\phi(t+1)-\phi(t)}_\alpha$ vanishes from the momentum conservation. 
%
\begin{figure}
\centering
\includegraphics[height=16cm, width=16cm, clip]{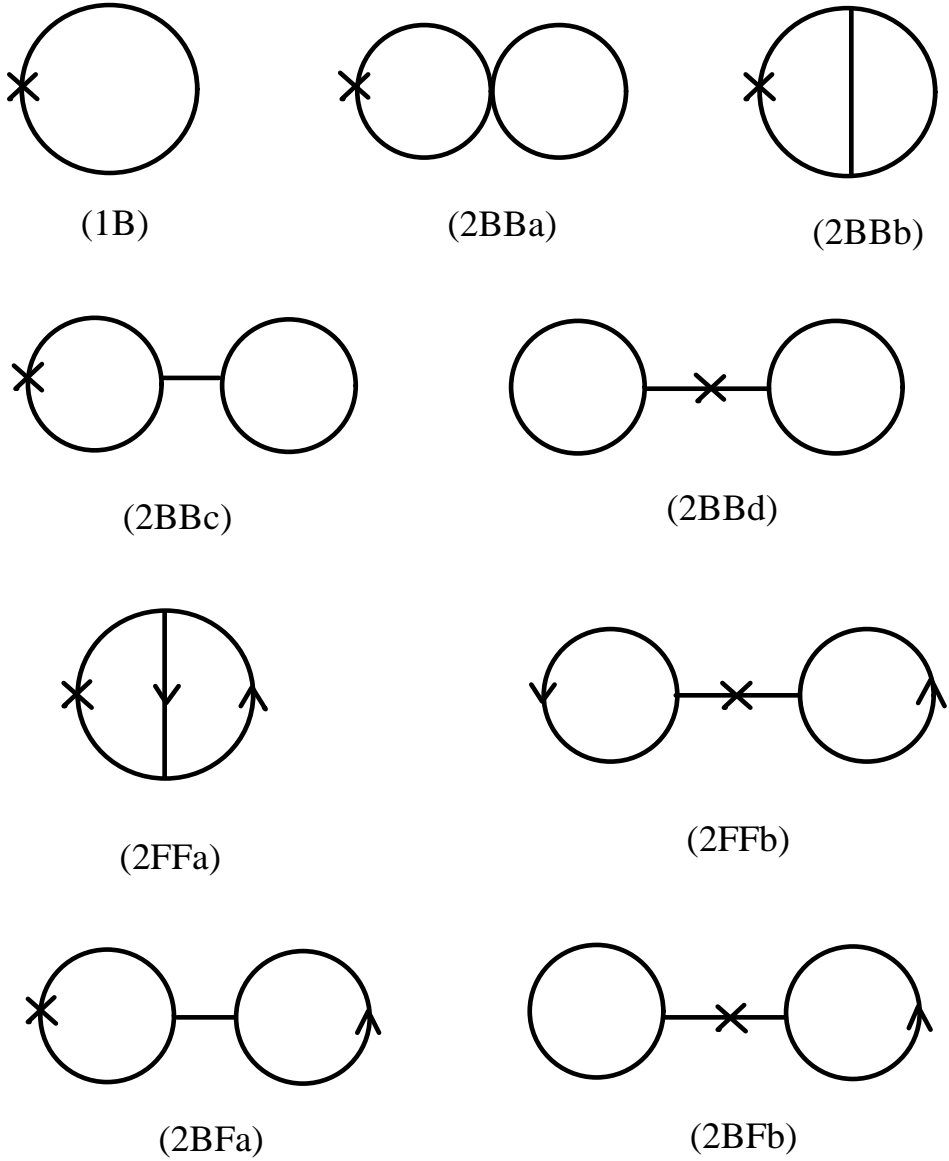}
\caption{One- and two-loop diagrams. The crosses represent the insertion of the operator $-ig\phi(t)^2$. 
The solid lines with (without) arrows mean the fermion (boson) propagators. 
(1B) is the one-loop diagram, and the other eight are the two-loop diagrams. 
The diagrams with the name ``FF'' (``BB'') are constructed by using the interaction vertices 
$V_{3F}$ twice ($V_4$ once or $V_{3B}$ twice), 
and those with ``BF'' are by using each of $V_{3B}$ and $V_{3F}$ once.}
\label{fig:2loop}
\end{figure}
%
For $-i\vev{g\phi(t)^2}_\alpha$, the one-loop contribution comes from the diagram (1B) in Fig.~\ref{fig:2loop}, 
which consists only of a boson line independent of $\alpha$. 
Also, the two-loop diagrams (2BBa), (2BBb), (2BBc) and (2BBd) do not contain fermion lines. 
The relevant diagrams for the IR divergence at the two-loop order are the last four (2FFa), (2FFb), (2BFa) and (2BFb), 
which are evaluated as 
\bea
{\rm (2FFa)} & = & i\frac{4g^3}{T^2} \sum_{m,k = -(T-1)/2}^{(T-1)/2} 
\left(\frac{1}{4\sin^2\left(\frac{\pi m}{T}\right)+M^2}\right)^2\frac{1}{e^{i(2\pi k +\alpha)/T}-1} 
\frac{1}{e^{i(2\pi (m+k) +\alpha)/T}-1},
 \nn \\
{\rm (2FFb)} & = & -i\frac{4g^3}{T^2} \frac{1}{M^4} 
\left(\sum_{m = -(T-1)/2}^{(T-1)/2} \frac{1}{e^{i(2\pi m +\alpha)/T}-1} \right)^2, \nn \\
{\rm (2BFa)} & = & -i\frac{4g^3}{T^2} \frac{1}{M^2} \sum_{m = -(T-1)/2}^{(T-1)/2} 
\left(1-\frac{M^2}{4\sin^2\left(\frac{\pi m}{T}\right)+M^2}\right)\frac{1}{4\sin^2\left(\frac{\pi m}{T}\right)+M^2} \nn \\
 & & \hspace{1.6cm} \times \sum_{k = -(T-1)/2}^{(T-1)/2} \frac{1}{e^{i(2\pi k +\alpha)/T}-1}, \nn \\
{\rm (2BFb)} & = &  -i\frac{4g^3}{T^2} \frac{1}{M^4} \sum_{m = -(T-1)/2}^{(T-1)/2} 
\left(1-\frac{M^2}{4\sin^2\left(\frac{\pi m}{T}\right)+M^2}\right) 
\sum_{k = -(T-1)/2}^{(T-1)/2} \frac{1}{e^{i(2\pi k +\alpha)/T}-1}. \nn \\
 & & 
\eea
Each diagram is singular as $\alpha \to 0$ due to the fermion zero-mode, 
however it is remarkable that the sum of them vanishes: 
\bea
& & {\rm (2FFa)} +  {\rm (2FFb)} +  {\rm (2BFa)} +  {\rm (2BFb)} \nn \\
& & = -i\frac{4g^3}{T^2} \frac{1}{M^4}\sum_{m=1}^{T-1}
\left[1-\left(\frac{M^2}{4\sin^2\left(\frac{\pi m}{T}\right)+M^2}\right)^2\right] F(m)
\eea
with 
\bea
F(m) & \equiv & \sum_{k=1}^T\left(1+\frac{1}{e^{i(2\pi (m+k) +\alpha)/T}-1}\right)
\frac{1}{e^{i(2\pi k +\alpha)/T}-1} \nn \\
& = & \sum_{k=1}^T\frac{1}{e^{i(2\pi k +\alpha)/T}-1}
\left[1-\frac{e^{-i(2\pi k +\alpha)/T}}{1-e^{i2\pi m/T}}
-\frac{e^{-i(2\pi k +\alpha)/T}}{1-e^{-i2\pi m/T}} \right] \nn \\
 & = & \sum_{k=1}^T e^{-i(2\pi k+\alpha)/T} = 0. 
\eea
Thus, the two-loop contribution turns out to have no $\alpha$-dependence, 
and the quantum corrections come only from 
the boson loops which are IR finite, that is consistent with (\ref{dQM_Bvev}). 
Since the classical value $-ig\mu^2=-i\frac{M^2}{2g}$ is regarded as $\cO(g^{-1})$, and 
$\ell$-loop contributions are of the order $\cO(g^{2\ell-1})$, 
the quantum corrections can not be comparable to the classical value in the perturbation 
theory. 
Thus, the conclusion of the SUSY breaking based on the classical value continues to be correct 
even at the quantum level.

\section{Overview of SUSY matrix models with TBC} 
\label{sec:MM}
\setcounter{equation}{0}
In this section, we define a matrix model analog of the discretized SUSY quantum mechanics 
with the TBC as discussed in the previous section,  
and give an overview of its general features in the large-$N$ limit. 

Let us begin with a matrix-model analog of \eqref{dQM_S} 
\be
S = Q \sum_{t=1}^T N\tr \left[\bar{\psi}(t) \left\{\frac12 B(t) +i\left(\phi(t+1) -\phi(t) + W'(\phi(t))\right)
\right\} \right]
\ee
with the $Q$-supersymmetry nilpotent ($Q^2=0$): 
\bea
Q\phi(t)=\psi(t), & & Q\psi(t)=0, \nn \\
Q\bar{\psi}(t) =-iB(t), & & QB(t)=0. 
\eea
All the variables are $N\times N$ hermitian matrices and 
the derivative of the superpotential is a polynomial of $\phi$ given as 
\be
W'(\phi(t)) =  \sum_pg_p \,\phi(t)^p.
\ee
Then, the action is 
\bea
S & = & \sum_{t=1}^T N\tr \left[\frac12 B(t)^2 +iB(t)\left\{ \phi(t+1)-\phi(t)+W'(\phi(t)) \right\} \right. 
\nn \\
 & & \hspace{2cm} \left. \frac{}{} +\bar{\psi}(t) \left(\psi(t+1) -\psi(t)\right) 
         +\sum_pg_p\sum_{k=0}^{p-1} \bar{\psi}(t)\phi(t)^k\psi(t)\phi(t)^{p-1-k}\right]. \nn \\
\eea
We will focus on the simplest case $T=1$ with the TBC~(\ref{TBC_alpha}) again~\footnote{
Similarly to section~\ref{sec:discrete_phi4}, for $T=1$ with the PBC, the action is supersymmetric under $Q$ 
as well as under $\bar{Q}$ 
for a general superpotential, and it can be expressed as 
\be
S = Q\bar{Q}\, \tr\left(\frac12\bar{\psi}\psi+W(\phi)\right).
\ee
The $Q, \bar{Q}$ transformations are of the same form as in (\ref{QSUSY}) and (\ref{Q_bar_T=1}), 
and nilpotent ($Q^2 = \bar{Q}^2 = \{Q, \bar{Q}\}=0$).}. 
Under the action 
\bea
S_\alpha & = & N \tr \left[\frac12 B^2 +iBW'(\phi) 
+\bar{\psi}\left(e^{i\alpha}-1\right)\psi +\sum_pg_p\sum_{k=0}^{p-1} \bar{\psi}\phi^k\psi\phi^{p-1-k}\right], 
\label{S_TBC}
\eea
the partition function is defined by
\be
Z_\alpha  \equiv    \left(-1\right)^{N^2}
\int d^{N^2}B \,d^{N^2}\phi \,d^{N^2}\psi \,d^{N^2}\bar{\psi}\, e^{-S_\alpha}.  
\label{Z_TBC}
\ee
For convenience, we fix the normalization of the measure as 
\be
\int d^{N^2}\phi \, e^{-N\tr \,(\frac12 \phi^2)} = \int d^{N^2}B \, e^{-N\tr \,(\frac12 B^2)} = 1, \qquad 
(-1)^{N^2} \int d^{N^2}\psi \,d^{N^2}\bar{\psi}\, e^{-N\tr \,(\bar{\psi}\psi)}=1. 
\label{normalization}
\ee

In order to construct SUSY matrix models 
which exhibit spontaneous SUSY breaking in the large-$N$ limit,  
we introduce a small positive constant $\epsilon$ and consider the situation that 
the derivative of the superpotential $W'(\phi)$ depends on $\phi$ 
only through the combination of $\epsilon\phi$, i.e. 
\be
W'(\phi) = h(\epsilon\phi) = \sum_p g_p (\epsilon\phi)^p 
\ee
with $g_p$ constants of order 1. 
Later 
we will discuss spontaneous SUSY breaking in the case that $\epsilon$ tends to 0 in the large-$N$ limit. 
(For example, $\epsilon$ scales as $\cO(N^{-u})$ with $u>0$.)   
Since $\epsilon^{-1}$ corresponds to the length scale of $\phi$ in which $W'(\phi)$
varies by an amount of $\cO(1)$, 
the potential $W'(\phi)$ becomes more and more flatter or slowly varying 
as $N\to \infty$~\footnote{In the case of large $T$, 
we can consider the continuum limit where the lattice spacing of the discretized time $a\to 0$ with $aT$ fixed. 
It would be interesting if $\epsilon$ can be also related to $a$ in the large-$N$ limit. 
It is similar to the situation of the double scaling limit in the matrix models for two-dimensional 
quantum gravity~\cite{Brezin:1990rb,Douglas:1989ve,Gross:1989aw}.}.  

Instead of taking the limits $N\to \infty$ and $\epsilon\to 0$ simultaneously, 
we can consider $\epsilon$ as independent of $N$, and can take the limits separately. 
Note that the external field $\alpha$ is always turned off at the final step   
because it is for detecting spontaneous SUSY breaking of the system obtained after the large-$N$ limit 
and the $\epsilon\rightarrow 0$ limit. 
In our model, it is easy to check that irrespective of the order of the limits $N\to \infty$ and $\epsilon\to 0$, 
we will have the same large-$N$ saddle point equation at the final step $\alpha\to 0$, 
and thus we obtain the same result. Therefore, we do not have to care about the order of these limits. 

After integrating $\psi, \bar{\psi}, B$ and rescaling as $\phi \to \epsilon^{-1}\phi$, 
the partition function becomes
\bea
Z_\alpha & = & \frac{1}{\epsilon^{N^2}}\, Z'_\alpha, \nn \\
Z_\alpha' & \equiv & \int d^{N^2}\phi \, \det \left[\left(e^{i\alpha}-1\right)\id \otimes\id 
+\epsilon\sum_pg_p\sum_{k=0}^{p-1}\phi^k\otimes\phi^{p-1-k}\right]
e^{-N\tr \left[\frac{1}{2}h(\phi)^2\right]} . \nn \\ 
\label{Z_MM}
\eea
The expectation values of $\frac1N\tr \,B^n$ ($n=1,2, \cdots$) are expressed as 
\bea
\vev{\frac{1}{N}\tr \,B^n}_\alpha & \equiv & \frac{1}{Z_\alpha} (-1)^{N^2}
\int d^{N^2}B \,d^{N^2}\phi \,d^{N^2}\psi \,d^{N^2}\bar{\psi} \,\left(\frac{1}{N} \tr \,B^n\right)
\, e^{-S_\alpha} \nn \\
 & = & \frac{1}{Z'_\alpha}\int d^{N^2}B \,d^{N^2}\phi \,\left(\frac{1}{N}\tr \,(B-ih(\phi))^n\right)
e^{-N\tr\left[\frac12B^2\right]} \nn \\
  & & \times \det \left[\left(e^{i\alpha}-1\right)\id \otimes\id 
+\epsilon\sum_pg_p\sum_{k=0}^{p-1}\phi^k\otimes\phi^{p-1-k}\right]
e^{-N\tr \left[\frac{1}{2}h(\phi)^2\right]} .  \nn\\
\label{B_MM} 
\eea

To figure out how the large-$N$ and slowly varying limit causes spontaneous SUSY breaking, 
we first regard $\epsilon$ as independent of $N$, 
and consider the case of large $N$ but finite $\epsilon$ in section~\ref{subsec:MM_finite_ep}.  
Then, the large-$N$ and slowly varying limit is discussed in sections~\ref{sec:MM_TBC} 
and \ref{sec:MM_deformation}. 

\subsection{Large-$N$ solutions for finite $\epsilon$}
\label{subsec:MM_finite_ep}
We consider large-$N$ solutions for the system (\ref{Z_MM}), (\ref{B_MM}) with 
$\epsilon$ kept finite. 
Here, we do not take the $\epsilon \to 0$ limit, and 
the external field $\alpha$ is turned off after the large-$N$ limit. 
(\ref{Z_MM}) can be expressed as integrals with respect to the eigenvalues of $\phi$ 
($\lambda_i$ ($i=1, \cdots, N$)):
\bea
Z'_\alpha & = &  \tilde{C}_N\int \left(\prod_{i=1}^N d\lambda_i\right)\, \Delta(\lambda)^2 \, 
\prod_{i,j=1}^N\left[e^{i\alpha}-1 + \epsilon \sum_pg_p\sum_{k=0}^{p-1}\lambda_i^k\lambda_j^{p-1-k}\right] \nn \\ 
 & & \hspace{3cm}\times e^{-N\sum_{i=1}^N \frac12h(\lambda_i)^2}
 \label{Z'_MM_lambda}
\eea
with $\tilde{C}_N$ a numerical constant and $\Delta(\lambda) \equiv \prod_{i>j} (\lambda_i-\lambda_j)$. 
In the large-$N$ limit, the integrals can be evaluated at the saddle point configuration as 
\bea
Z'_{\alpha} & = & C_N \exp\left[-N^2\left(F_{\alpha} +\cO(N^{-2})\right) \right], \label{Z_alpha_finite_ep} \nn \\
F_{\alpha} & = & -\int dx\, dy \rho(x)\rho(y)\,{\rm P}\ln|x-y| \nn \\
 & & 
-\int dx \,dy \rho(x)\rho(y)\,\ln\left(e^{i\alpha}-1 +\epsilon\sum_pg_p\sum_{k=0}^{p-1}x^ky^{p-1-k}\right) \nn \\
 & & +\int dx \,\frac12h(x)^2\rho(x)
\label{F_alpha_finite_ep}
\eea
with $C_N=\exp\left[\frac34N^2 +\cO(N^0) \right]$ obtained in (\ref{C_N}) in appendix~\ref{app:gaussian}. 
The eigenvalue distribution $\rho(x)=\frac{1}{N}\sum_{i=1}^N \delta(x-\lambda_i)$ satisfies the large-$N$ 
saddle point equation: 
\be
\int dy\,\rho(y)\,{\rm P}\frac{1}{x-y} 
+\int dy\,\rho(y)\,\frac{\epsilon\sum_p g_p \sum_{k=1}^{p-1}kx^{k-1}y^{p-1-k}}{e^{i\alpha}-1
+\epsilon\sum_p g_p \sum_{k=0}^{p-1}x^{k}y^{p-1-k}} = \frac12 h(x)h'(x). 
\label{saddle_pt_finite_ep0}
\ee
Finally, after the limit $\alpha\to 0$, we obtain  
\be
 \int dy\,\rho(y)\,{\rm P}\frac{1}{x-y}
+\int dy\,\rho(y)\,
\frac{\sum_p g_p \sum_{k=1}^{p-1}kx^{k-1}y^{p-1-k}}{\sum_p g_p \sum_{k=0}^{p-1}x^{k}y^{p-1-k}} 
= \frac12 h(x)h'(x). 
\label{saddle_pt_finite_ep1}
\ee
Note that the $\epsilon$-dependence disappears in the equation.

In terms of $\rho(x)$ which solves (\ref{saddle_pt_finite_ep1}), 
the expectation values of $\frac1N\tr \,B^n$ ($n=1,2,\cdots$) are given by 
\bea
\lim_{\alpha\to 0}\left(\lim_{N\to \infty} \vev{\frac1N\tr \,B}_{\alpha} \right) & = & 
-i\int dx \rho(x)h(x),   \nn \\
\lim_{\alpha\to 0}\left(\lim_{N\to \infty} \vev{\frac1N\tr \,B^2}_{\alpha} \right) & = & 
1-\int dx \rho(x)h(x)^2, \nn \\
\lim_{\alpha\to 0}\left(\lim_{N\to \infty} \vev{\frac{1}{N}\tr \,B^3}_{\alpha} \right) & = & 
-3i\int dx \, \rho(x) h(x) 
+ i\int dx \, \rho(x) h(x)^3 , \nn \\
\lim_{\alpha\to 0}\left(\lim_{N\to \infty} \vev{\frac{1}{N}\tr \,B^4}_{\alpha} \right) & = & 
2-4\int dx \, \rho(x) h(x)^2 
                -2\left[\int dx \,\rho(x) h(x)\right]^2 \nn \\
    & & + \int dx \, \rho(x) h(x)^4 ,\nn \\
\vdots & & .\label{Bvev_MM_finite_ep}   
\eea
 
In section~\ref{subsec:phi4SUSY}, we explicitly solve the saddle point equation for 
the quadratic $W'(\phi)$: $W'(\phi) = h(\epsilon\phi)=(\epsilon\phi)^2+\mu^2$ 
by the technique of the $O(n)$ model on a random surface 
with $n=-2$~\cite{Kostov:1988fy,Kostov:1992pn,Eynard:1995nv}.

Here it is important to recall that, when both of $\epsilon$ and $N$ are finite, 
basically we will obtain the same result as in the discretized SUSY quantum mechanics. 
Namely, when $W'(\phi)$ is a polynomial of odd (even) degree, 
the SUSY is preserved (broken).   
However, a new feature can arise 
in the large-$N$ limit even if $\epsilon$ is kept finite. 
It is analogous to the infinite volume limit of SUSY field theory~\cite{Witten:1982df}.  
There, in the case of the double-well potential, the SUSY breaking is triggered by instanton effects in finite volume,  
however the instantons are suppressed in the large volume limit, and the SUSY becomes restored. 
We will find in section \ref{subsec:phi4SUSY} that the large-$N$ limit plays the same role 
as the infinite volume limit and restores the SUSY for the same potential. 

\subsection{Large-$N$ SUSY breaking under slowly varying potential}
\label{sec:MM_TBC}
In the large-$N$ and slowly varying limit, 
$N\to \infty$ together with $\epsilon \to 0$, 
we can discard the $\epsilon$-dependent terms in the determinant factor 
in (\ref{Z_MM}). Then, the system is essentially reduced to the ordinary one-matrix model 
with respect to $\phi$. 
In terms of the eigenvalue distribution $\rho(x)$, 
the partition function~$Z'_\alpha$ is given in the large-$N$ limit as 
\bea
Z'_\alpha & = & \left(e^{i\alpha}-1\right)^{N^2} C_N \exp\left[-N^2\left(F_0 +\cO(N^{-2})\right)\right],\nn \\
F_0 &=& -\int dx\, dy \rho(x)\rho(y)\,{\rm P}\ln|x-y| +\int dx \,\frac12 h(x)^2\rho(x) 
\label{F_0}
\eea
with $\rho(x)$ satisfying the saddle point equation
\be
\int dy \,\rho(y)\,{\rm P}\frac{1}{x-y} = \frac12 h(x)h^{\prime}(x). 
\label{SPE}
\ee
Note that the equation does not depend on $\alpha$. 
It can be solved in the standard method \cite{Brezin:1977sv} 
by assuming a finite support for $\rho(x)$. 
For example, given such a solution $\rho(x)$ with the support $x\in [a, b]$, 
\eqref{F_0} can be expressed by integrating \eqref{SPE} as 
\be
F_{0}=-\int dx\rho(x)\ln|b-x|+\int dx\frac14h(x)^2\rho(x)+\frac14h(b)^2.
\label{F_MM}
\ee 
As for the expectation values of $\frac1N\tr \,B^n$ ($n=1,2,\cdots$), 
we obtain 
\bea
\vev{\frac1N\tr \,B}_\alpha & = & -i\int dx\rho(x)h(x), \nn \\
\vev{\frac1N\tr \,B^2}_\alpha & = & 1-\int dx\rho(x)h(x)^2, \nn \\
\vev{\frac{1}{N}\tr \,B^3}_{\alpha} & = & 
-3i\int dx \, \rho(x) h(x) 
+ i\int dx \, \rho(x) h(x)^3 , \nn \\
\vev{\frac{1}{N}\tr \,B^4}_{\alpha} & = & 
2-4\int dx \, \rho(x) h(x)^2 
                -2\left[\int dx \,\rho(x) h(x)\right]^2 \nn \\
    & & + \int dx \, \rho(x) h(x)^4 ,\nn \\
\vdots & &  ,
\label{Bvev_alpha}
\eea
which do not depend on $\alpha$ similarly to the situation seen in the (discretized) SUSY quantum mechanics 
in section~\ref{sec:SQM}.

Also, 
\be
\vev{\frac{1}{N}\tr\left((\epsilon\phi)^n\right)}_\alpha= \int dx \,x^n \rho(x) \qquad (n=1,2,\cdots)
\ee
do not depend on $\alpha$ and exhibit no singular behavior as $\alpha\to 0$. 
It is different from the situation of the (discretized) quantum mechanics, 
where $\vev{\phi}_\alpha$ brows up in the limit $\alpha\to 0$ 
as in (\ref{phi_vev_alpha_QM}) and (\ref{phi_vev_alpha_dQM}). 
It implies that the large-$N$ limit suppresses the quantum tunneling to make the 
superselection rule work in the matrix model. 
Although $\vev{\frac{1}{N}\tr\left(\phi^n\right)}_\alpha$ grows as $\epsilon^{-n}$ 
in the slowly varying limit, 
any singular behavior does not appear for the ``renormalized'' variable $\epsilon\phi$. 


Now it is clear from \eqref{Z_MM} and \eqref{B_MM} how SUSY breaking happens 
in the $\epsilon\rightarrow 0$ limit. 
Since the large-$N$ and slowly varying limit amounts to dropping the fermion interactions, 
there is generically no reason to expect that the SUSY persists after turning off $\alpha$.  
It will be so regardless of the form of $W'(\phi)$, except the Gaussian case 
$W'(\phi)=h(\epsilon\phi)=\epsilon\phi$ 
where the fermion determinant is constant.
In the following sections, for some concrete examples of $W'(\phi)$, 
we compute the expectation values of $\frac1N\tr \, B^n$ ($n=1,2,\cdots$) from which we observe 
spontaneous SUSY breaking, and see that this is indeed the case.   
In particular, it is quite interesting that SUSY is broken 
even in the case of $W'(\phi)$ with odd degree, 
for which the SUSY is preserved in the (discretized) SUSY quantum mechanics. 
Since the finite $N$ case will lead to the same conclusion as in the discretized quantum mechanics, 
our model provides some new 
examples of spontaneous SUSY breaking in that the $\alpha\rightarrow 0$ limit 
and the large-$N$ limit do not commute. 
It is more analogous to spontaneous breaking of ordinary 
bosonic symmetry. 

As seen from \eqref{Z_MM} and \eqref{B_MM}, without the twist $\alpha$,  
we cannot take the $\epsilon\rightarrow 0$ limit in a well-defined manner. In a sense, this is as usual: 
in order to discuss spontaneous symmetry breaking, introduction of the external field 
is essential, 
because it lifts degeneracy of vacua and makes the notion of the vacuum expectation value well-defined 
as stressed repeatedly in the previous section.

\subsection{Another deformation}
\label{sec:MM_deformation}
So far, we introduced the external field $\alpha$ 
to see the spontaneous SUSY breaking. 
It can be regarded as a deformation of the model with inserting the operator, 
which changes the boundary condition from the PBC to the TBC by the twist $\alpha$. 
The deformation slightly breaks the SUSY of the original model. 
Then, what we saw in section~\ref{sec:MM_TBC} is interpreted as a statement that, 
if the limit $\alpha\to 0$ is taken after the large-$N$ and slowly varying limit, 
the SUSY breaking effect remains even after turning off the deformation $\alpha$.   

To see how the model is sensitive to the deformation, 
we here consider another deformation corresponding to a specific twisted boundary condition, 
where the twist also depends on the variable $\phi$: 
\be
\phi(2) = \phi(1), \qquad \psi(2) = e^{i\alpha\frac{1}{N}\tr V(\epsilon\phi(1))}\psi(1),
\label{deformation}
\ee
where $V(\phi)$ is a function of $\phi$. 
Then, the action reads  
\be
S_\alpha = N \tr \left[\frac12 B^2 +iBW'(\phi) 
+\bar{\psi}\left(e^{i\alpha\frac{1}{N}\tr V(\epsilon\phi)}-1\right)\psi 
+\sum_p\epsilon^pg_p\sum_{k=0}^{p-1}\bar{\psi}\phi^k\psi\phi^{p-1-k}\right]. 
\ee

Note that the twist depends on $\phi$ through the combination $\epsilon\phi$, 
which is the same as the $\phi$-dependence of the potential $W'(\phi) =h(\epsilon\phi)$.   
We consider such a setting in order to make well-defined 
the limit $\alpha \to 0$ after the large-$N$ and slowly varying limit. 
The $\phi$-dependent twist could also be interpreted as some operator inserted at 
the boundary.    

Similarly, after the rescaling $\phi \to \epsilon^{-1}\phi$, 
the partition function 
and the expectation values of $\frac{1}{N}\tr \,B^n$ ($n=1,2,\cdots$) become 
\bea
Z'_\alpha & \equiv & \int d^{N^2}\phi \, 
\det \left[\left(e^{i\alpha\frac{1}{N}\tr \,V(\phi)}-1\right)\id \otimes\id 
+\epsilon\sum_pg_p\sum_{k=0}^{p-1}\phi^k\otimes\phi^{p-1-k}\right]
\nn \\
&& \hspace{1.5cm} \times e^{-N\tr \left[\frac{1}{2}h(\phi)^2\right]} , \nn \\
 & &   \\
\vev{\frac{1}{N}\tr \,B^n}_\alpha 
 & = & \frac{1}{Z'_\alpha}\int d^{N^2}B \,d^{N^2}\phi \,\left(\frac{1}{N}\tr \,(B-ih(\phi))^n\right)
e^{-N\tr\left[\frac12B^2\right]} \nn \\
& & \hspace{2cm} \times  
\det \left[\left(e^{i\alpha\frac{1}{N}\tr \,V(\phi)}-1\right)\id \otimes\id +\epsilon\sum_pg_p\sum_{k=0}^{p-1}\phi^k\otimes\phi^{p-1-k}\right]
\nn \\
 & & 
\hspace{2cm}\times e^{-N\tr \left[\frac{1}{2}h(\phi)^2\right]} .  
\eea

In the large-$N$ and slowly varying limit, 
the $\epsilon$-dependent part in the determinant factor can be neglected. 
Then, the partition function $Z'_\alpha$ at a large-$N$ saddle point 
is expressed as 
\bea
Z'_\alpha & = & C_N \exp\left[-N^2\left(F_\alpha +\cO(N^{-2})\right)\right], \nn \\ 
F_\alpha &=& -\int dx\, dy \rho(x)\rho(y)\,{\rm P} \ln|x-y| 
-\ln\left(e^{i\alpha\int dx \,V(x)\rho(x)}-1\right) 
+\int dx \,\frac12 h(x)^2 \rho(x),  \nn \\
 & & 
\label{F_deform}
\eea
where the eigenvalue distribution $\rho(x)$ is determined by 
\be
\int dy\,\rho(y) \,{\rm P}\frac{1}{x-y} + \frac{\frac12i\alpha V'(x)}{1-e^{-i\alpha c}} 
-\frac12h(x)h^{\prime}(x)=0
\label{eq_deform0}
\ee
with $c \equiv \int dy\, V(y)\rho(y)$. 
Note that the second term remains nonzero and contributes as a part of 
the potential~\footnote{This would be a peculiarity of our model with $T=1$. } 
even after the limit $\alpha\to 0$: 
\be
\int dy\,\rho(y) {\rm P}\frac{1}{x-y} = \frac12h(x)h^{\prime}(x)-\frac{1}{2c}V'(x).
\label{eq_deform} 
\ee
Therefore, our problem is reduced to that of the one-matrix model. 
In this case, (\ref{eq_deform}) should be solved consistently with $c=\int dy V(y)\rho(y)$. 
In the following sections, for various $W'(\phi)$ and $V(\epsilon\phi)$, 
we present the solutions.  
As a relevant quantity for the planar free energy after the limit $\alpha\to 0$, we will consider 
\bea
F_0 & \equiv & \lim_{\alpha\to 0} \left(F_\alpha +\ln(i\alpha)\right) \nn \\
   & = & -\ln c -\int dx\, dy \rho(x)\rho(y)\,{\rm P} \ln|x-y| 
+\int dx \,\frac12 h(x)^2 \rho(x). 
\label{phi4_MM_F_0_deform}
\eea
Given such a solution $\rho(x)$, for example with a support $x\in[a, b]$, 
we can express (\ref{phi4_MM_F_0_deform}) as 
\be
F_0 = -\ln c -\int dx \,\rho(x)\ln|b-x| +\frac{1}{2c}(c-V(b))
+\int dx \,\rho(x)\frac14h(x)^2+\frac14h(b)^2.
\ee
The expectation values $\vev{\frac1N\tr\,B^n}_\alpha$ ($n=1,2,\cdots$) after $\alpha$ turned off 
are expressed in the same form as (\ref{Bvev_alpha}). 
Note that it is finite and well-defined in the limit $\alpha \to 0$. 
For the discretized quantum mechanics with the slowly varying potential $W'(\phi) =h(\epsilon\phi)$ 
of odd degree and 
the $\phi$-dependent twists $\alpha V(\epsilon\phi)$, 
it will be observed that some of $\vev{B^n}_\alpha$ are singular as $\alpha\to 0$ after the slowly varying limit 
$\epsilon\to 0$. Since this kind of pathological behavior does not appear in the case of 
the large-$N$ matrix model, it shows further importance of the large-$N$ limit to construct the models.  

For the same reason as in section~\ref{sec:MM_TBC}, 
we can say that spontaneous SUSY breaking happens for any $W'(\phi)$ 
in the case of the $\phi$-dependent twist \eqref{deformation}. 
Since, differently 
from the constant twist \eqref{TBC_alpha}, 
the twist $V(\phi)$ now takes part in the saddle point equation~\eqref{eq_deform}, 
the twist $V(\epsilon\phi)$ causes spontaneous SUSY breaking even for 
the Gaussian case~$W'(\phi)=h(\epsilon\phi)=\epsilon\phi$, 
as we will see in section~\ref{sec:gaussian_MM}. 

\section{Large-$N$ $\phi^4$ SUSY matrix model} 
\label{sec:phi4_MM}
\setcounter{equation}{0}
In this section, we concretely examine the case with 
\be
W'(\phi) = h(\epsilon\phi) = \left(\epsilon\phi\right)^2 +\mu^2 
\label{phi4_MM_W'}
\ee
in the setting of section~\ref{sec:MM}.

\subsection{Large-$N$ solutions for finite $\epsilon$ }
\label{subsec:phi4SUSY}
Here, we will see that the system (\ref{Z_MM}) with (\ref{phi4_MM_W'}) 
for general finite $\epsilon$ has large-$N$ saddle points, which 
is expected to preserve SUSY for (some region of) $\mu^2<0$. 
For finite $N$, on the other hand, the system will take place spontaneous SUSY breaking, 
as we saw explicitly for the $N=1$ case in section~\ref{sec:discrete_phi4}.    
Thus, this presents an example of matrix models, 
where the SUSY is broken for finite $N$ but restores in the large-$N$ limit\footnote{Such an example 
for a large-$N$ vector model has been presented in Ref.~\cite{Affleck:1983pn}.}. 

The partition function reads 
\bea
Z_\alpha & = & \frac{1}{\epsilon^{N^2}}Z'_\alpha, \nn \\
Z'_{\alpha} & = & \tilde{C}_N\int \left(\prod_{i=1}^N d\lambda_i\right)\, \Delta(\lambda)^2 \, 
\prod_{i,j=1}^N\left[e^{i\alpha}-1 +\epsilon (\lambda_i+\lambda_j)\right] \, e^{-N\sum_{i=1}^N \frac12\left(\lambda_i^2+\mu^2\right)^2}, 
\label{phi4MM_Z'_lambda}
\eea
and the saddle point equation (\ref{saddle_pt_finite_ep1}) becomes 
\be
\int dy \, \rho(y) {\rm P}\frac{1}{x-y} + \int dy \, \frac{\rho(y)}{x+y} = x^3 +\mu^2 x . 
\label{phi4_MM_saddle_pt_finite_ep}
\ee
Let us consider the case $\mu^2 < 0$, where the shape of the potential is 
a double-well $\frac12 \left(x^2+\mu^2\right)^2$.

\subsubsection{One-cut solution}
\label{sec:phi4_one-cut}
First, we find a solution corresponding to all the eigenvalues located around 
one of the minima $\lambda= +\sqrt{-\mu^2}$. 
Assuming the support of $\rho(x)$ as $x\in [a,b]$ with $0<a<b$, 
the equation (\ref{phi4_MM_saddle_pt_finite_ep}) is valid for $x\in [a, b]$. 

Following \cite{Brezin:1977sv}, we introduce a holomorphic function 
\be
F(z)\equiv \int_a^b dy \frac{\rho(y)}{z-y}, 
\ee
which satisfies the following properties: 
\begin{enumerate}
\item
$F(z)$ is analytic in $z\in {\bf C}$ except the cut $[a,b]$ .
\item 
$F(z)$ is real on $z\in {\bf R}$ outside the cut. 
\item
For $z\sim \infty$, \\
$F(z) = \frac{1}{z} + \cO\left(\frac{1}{z^2}\right)$.
\item
For $x\in [a,b]$, \\
$F(x\pm i0) = F(-x) + x^3 +\mu^2 x \mp i\pi \rho(x)$.
\end{enumerate}

Note that, if we consider the combination~\cite{Eynard:1995nv}
\be
F_-(z) \equiv \frac12\left(F(z)-F(-z)\right),  
\ee
then the properties of $F_-(z)$ are 
\begin{enumerate}
\item
$F_-(z)$ is analytic in $z\in {\bf C}$ except the two cuts $[a,b]$ and $[-b,-a]$. 
\item
$F_-(z)$ is odd ($F_-(-z) = -F_-(z)$), and real on $z\in {\bf R}$ outside the cuts. 
\item
For $z\sim \infty$, \\
$F_-(z) = \frac{1}{z} + \cO\left(\frac{1}{z^3}\right)$.
\item
For $x\in [a,b]$, \\
$
F_-(x\pm i0) = \frac12 \left(x^3+\mu^2x\right) \mp i\frac{\pi}{2} \rho(x). 
$
\end{enumerate}

These properties are sufficient to fix the form of $F_-(z)$ as 
\be
F_-(z) = \frac12 \left(z^3+\mu^2 z\right) -\frac12 z\sqrt{(z^2-a^2)(z^2-b^2)}
\ee
with
\be
a^2 = -2-\mu^2, \qquad b^2 = 2-\mu^2. 
\label{ab_phi4_PBC}
\ee
Since $a^2$ should be positive, the solution is valid for $-\mu^2 > 2$. 
The eigenvalue distribution is obtained as 
\be
\rho(x) = \frac{x}{\pi} \sqrt{(x^2-a^2)(b^2-x^2)}.  
\label{rho_phi4_PBC}
\ee

Applying (\ref{rho_phi4_PBC}) to (\ref{Bvev_MM_finite_ep}), 
it is seen that all the expectation values of $\frac{1}{N}\tr \, B^n$ ($n=1,2,3,4$) vanish. 
Moreover, we can show 
\be
\lim_{\alpha\to 0}\left(\lim_{N\to \infty} \vev{\frac{1}{N}\tr \,B^n}_\alpha \right) =0 \qquad (n=1,2,\cdots)
\label{Bn=0}
\ee
by an inductive argument as in appendix~\ref{app:proof}.  

Also, the partition function can be calculated, and we see that it coincides with 
that of the Gaussian SUSY matrix model with $g_1>0$ obtained in appendix~\ref{app:gaussian}: 
\be
\lim_{\alpha\to 0}\left(\lim_{N\to \infty}\frac{1}{N^2}\ln Z_{\alpha}\right) = 
\lim_{N\to \infty} \frac{1}{N^2}\ln \left(\left. Z_G\right|_{g_1>0}\right) = 0.
\ee

These evidences convince us that the SUSY is restored at infinite $N$. 

\subsubsection{Two-cut solutions}  
Let us consider configurations that $\nu_+N$ eigenvalues are located around 
one minimum $\lambda=+\sqrt{-\mu^2}$ of the double-well, and 
the remaining $\nu_-N (= N-\nu_+N)$ eigenvalues are around the other minimum $\lambda=-\sqrt{-\mu^2}$. 
Since the fermion determinant in the partition function (\ref{phi4MM_Z'_lambda}) is 
\be
\prod_{i,j=1}^N\left[e^{i\alpha}-1+\epsilon(\lambda_i+\lambda_j)\right] 
= \prod_{i=1}^N \left[e^{i\alpha}-1+ 2\epsilon\lambda_i\right] 
\prod_{i> j} \left[e^{i\alpha}-1+\epsilon(\lambda_i + \lambda_j)\right]^2, 
\ee 
the configurations will contribute $(-1)^{\nu_-N}$ to the sign of the partition function 
for small $\alpha$.  
Thus, the sign is stable against variations around the configurations with $\nu_+$ or $\nu_-$ fixed. 
Therefore, in the large-$N$ limit the tunneling is suppressed and  
a smooth saddle point is expected to be found. 
Because such a saddle point will form a two-cut solution at large $N$, 
let us investigate two-cut solutions. 

First, we focus on the $Z_2$-symmetric two-cut solution with $\nu_+ =\nu_-=\frac12$, where 
the eigenvalue distribution is supposed to have a $Z_2$-symmetric support $\Omega=[-b, -a]\cup [a,b]$, 
and $\rho(-x)=\rho(x)$. 
The equation (\ref{phi4_MM_saddle_pt_finite_ep}) is valid 
for $x\in \Omega$. Due to the $Z_2$ symmetry, the holomorphic function 
$F(z) \equiv \int_\Omega dy\,\frac{\rho(y)}{z-y}$ has the same properties as $F_-(z)$ 
in section~\ref{sec:phi4_one-cut} except the property 4, which is now changed to 
\be
F(x\pm i0) = \frac12\left(x^3+\mu^2 x\right) \mp i\pi \rho(x) \qquad \mbox{for} \quad x\in \Omega. 
\ee
The solution is given by 
\bea
F(z) & = & \frac12\left(z^3+\mu^2 z\right) -\frac12 z\sqrt{(z^2-a^2)(z^2-b^2)}, \\
\rho(x) & = & \frac{1}{2\pi}|x|\sqrt{(x^2-a^2)(b^2-x^2)}, 
\eea
where $a$, $b$ coincide with the values of the one-cut solution (\ref{ab_phi4_PBC}). 
It is easy to see that, concerning $Z_2$-symmetric observables, the expectation values 
are same as the expectation values evaluated under the one-cut solution. 
In particular, 
the proof in appendix~\ref{app:proof} can be applied also here, to obtain 
\be
\lim_{\alpha\to 0}\left(\lim_{N\to \infty} \vev{\frac{1}{N}\tr \,B^n}_\alpha \right)=0 \qquad  (n=1,2,\cdots).
\ee 
The partition function with the sign factor dropped is the same as that evaluated by 
the one-cut solution:  
$\lim_{\alpha\to 0}\left(\lim_{N\to \infty} \frac{1}{N^2}\, \ln \left|Z_{\alpha}\right|\right)=0$. 

It is somewhat surprising that the end points of the cut $a$, $b$ and the absolute value of 
the partition function coincide with those for the one-cut solution, 
which is recognized as a new interesting feature of the supersymmetric models and 
can be never seen in the case of bosonic double-well matrix models.  
In bosonic double-well matrix models, the free energy of the $Z_2$-symmetric two-cut solution 
is lower than that of the one-cut solution, and the endpoints of the cuts are different between the 
two solutions~\cite{Cicuta:1986pu,Nishimura:2003gz}. 

Next, let us consider general $Z_2$-asymmetric two-cut solutions (i.e., general $\nu_\pm$). 
We can check that the following solution gives a large-$N$ saddle point: \\
The eigenvalue distribution $\rho(x)$ has the cut $\Omega=[-b,-a]\cup [a,b]$ with $a$, $b$ given by 
(\ref{ab_phi4_PBC}): 
\be
\rho(x) = \left\{\begin{array}{cl} \frac{\nu_+}{\pi}\,x\,\sqrt{(x^2-a^2)(b^2-x^2)} & \qquad (a<x<b) \\
                                   \frac{\nu_-}{\pi}\,|x|\,\sqrt{(x^2-a^2)(b^2-x^2)} & \qquad (-b<x<-a).
                 \end{array}\right. 
\label{general_two-cut_phi4_PBC}                 
\ee 
This is a general supersymmetric solution including the one-cut and $Z_2$-symmetric two-cut 
solutions. 
The expectation values of $Z_2$-even observables under this saddle point coincide those under the one-cut 
solution, and 
$\lim_{\alpha\to 0}\left(\lim_{N\to \infty} \vev{\frac{1}{N}\tr \,B^n}_\alpha \right) =0$ ($n=1,2,\cdots$), again. 
Also, for the partition function, 
$\lim_{\alpha\to 0}\left(\lim_{N\to \infty} \frac{1}{N^2}\, \ln \left|Z_{\alpha}\right|\right)=0$.   
Thus, we can conclude that the SUSY matrix model with the double-well potential has an infinitely 
many degenerate supersymmetric saddle points parametrized by $(\nu_+, \nu_-)$ at large $N$ 
for the case $\mu^2<-2$. 
It is totally different from the $N=1$ case discussed in section~\ref{subsub:T=1}, 
where the partition function with the PBC vanishes and the SUSY is broken.

For the general two-cut solution with ($\nu_+,\nu_-$), the partition function  
appears to be expressed as 
\be
\lim_{\alpha\to 0}\left(\lim_{N\to \infty}\frac{1}{N^2}\ln \left|Z_{\alpha}\right|\right) = 
\lim_{N\to \infty} \frac{1}{N^2}\ln \left| Z_{G, \nu_+}Z_{G, \nu_-}\right| = 0.
\ee 
Here, $Z_{G, \nu_\pm}$ are the partition functions of the Gaussian SUSY matrix models with the matrix size 
$\nu_\pm N$, 
whose superpotential is obtained by the Gaussian approximation of the original superpotential 
$W'(x)=h(\epsilon x) =(\epsilon x)^2+\mu^2$ 
around its minima $x=\pm \frac{1}{\epsilon}\sqrt{-\mu^2}$, respectively. 
Furthermore, since $Z_\alpha$ and $Z_{G, \nu_+}Z_{G, \nu_-}$ have the same sign factor $(-1)^{\nu_-N}$, 
$Z_\alpha$ can be evaluated by the product of the partition functions of 
the Gaussian SUSY matrix models $Z_{G, \nu_+}Z_{G, \nu_-}$ including the sign factor at large $N$. 
It is reasonable to attribute these remarkable features for general $\nu_{\pm}$ to the restoration 
of the SUSY. Details of this aspect will be reported in the next publication.

\subsection{Large-$N$ SUSY breaking under slowly varying potential}
\label{sec:phi4_MM_TBC}
Here, we show that the model with the same superpotential (\ref{phi4_MM_W'}) 
exhibits spontaneous SUSY breaking in the large-$N$ and slowly varying limit. 

The saddle point equation (\ref{SPE}) becomes
\be
\int dy \,\rho(y)\,{\rm P}\frac{1}{x-y} = x^3+\mu^2x. 
\ee

\paragraph{One-cut solution} 
Assuming the support of $\rho(x)$ as $x\in [-a,a]$ and using the standard method~\cite{Brezin:1977sv}, 
it can be solved as 
\bea
\rho(x) & = & \frac{1}{\pi}\left(x^2+\mu^2 +\frac{a^2}{2}\right)\sqrt{a^2-x^2}, \\ 
\label{phi4_MM_TBC_one-cut}
a^2  & =  & \frac23\left(-\mu^2 +\sqrt{\mu^4 +6}\right). 
\eea
{}From the requirement $\rho(x)\ge 0$, this solution is valid for $\mu^2\ge -\sqrt{2}$. 

Then, the expectation value of $\frac{1}{N}\tr \,B$ becomes independent of $\alpha$ at large $N$: 
\bea
\vev{\frac{1}{N}\tr \,B}_\alpha & = &  -i \int^a_{-a}\dd x \,(x^2+\mu^2)\rho(x) \nn \\
  & = & -i\frac18 a^2\left(a^2+2\mu^2\right)^2.
\eea
This is nonzero for $\mu^2>-\sqrt{2}$, meaning the SUSY is broken. 
It accidentally vanishes at $\mu^2=-\sqrt{2}$, 
but since 
\bea
\vev{\frac{1}{N}\tr \,B^2}_\alpha & = &  1- \int^a_{-a} dx \,(x^2+\mu^2)^2\rho(x) \nn \\
  & = & \frac{1}{2} \neq 0 \qquad \mbox{for} \quad \mu^2=-\sqrt{2}, 
\eea 
the SUSY breaking is seen also at $\mu^2=-\sqrt{2}$. 

As a quantity relevant to the planar free energy 
(after subtracted by the divergent constant as $\alpha\to 0$), 
we compute (\ref{F_0}): 
\be
F_0 = -\int dx \,dy\, \rho(x) \rho(y)\, {\rm P}\ln |x-y| 
+ \int dx \,\frac12 \left(x^2+\mu^2\right)^2 \rho(x). 
\ee
Plugging the one-cut solution (\ref{phi4_MM_TBC_one-cut}) into \eqref{F_MM}, we obtain 
\be
\left. F_0\right|_{\rm one-cut} = 
-\ln \frac{a}{2} +\frac38 +\frac{17}{36}\mu^4 +\left(\frac{5}{24}\mu^2 +\frac{1}{72}\mu^6\right) a^2. 
\label{F_onecut}
\ee 

\paragraph{Two-cut solution} 
For $\mu^2 <-\sqrt{2}$, the solution develops to have two cuts. 
Among various two-cut solutions, the most relevant 
is a $Z_2$-symmetric one~\cite{Cicuta:1986pu,Nishimura:2003gz}, where the eigenvalue distribution has two supports $[-b, -a]$ and $[a, b]$. 
The explicit form is  
\be
\rho(x) = \frac{1}{\pi} \,|x|\sqrt{(x^2-a^2)(b^2-x^2)}
\ee
with $a^2 = -\mu^2 -\sqrt{2}$, $b^2 = -\mu^2 +\sqrt{2}$. 

Interestingly, the free energy for this solution is 
turned out to be $\mu^2$-independent: 
\be
\left. F_0\right|_{\rm two-cut} = \frac38 +\frac14 \ln 2, 
\label{F_twocut}
\ee  
and the expectation values of $B$ are 
\be
\vev{\frac{1}{N}\tr \,B}_\alpha =0, \qquad \vev{\frac{1}{N}\tr \,B^2}_\alpha =\frac12.
\ee
The SUSY is broken also at the two-cut phase. 
The property of the free energy might be interpreted as some remnant of supersymmetric properties  
of the model. 

\paragraph{Phase transition}
Comparing $\left. F_0\right|_{\rm one-cut}$ and $\left. F_0\right|_{\rm two-cut}$ at the phase boundary 
$\mu^2 =-\sqrt{2}$, from \eqref{F_onecut} and \eqref{F_twocut} 
it is seen that $F_0$, $\frac{d}{d(\mu^2)}F_0$ and $\frac{d^2}{d(\mu^2)^2}F_0$ is continuous, but 
\be
\lim_{\mu^2\rightarrow -\sqrt{2}+0}\frac{d^3}{d(\mu^2)^3}F_0 = \frac{1}{\sqrt{2}}, ~~~~~
\lim_{\mu^2\rightarrow -\sqrt{2}-0}\frac{d^3}{d(\mu^2)^3}F_0=0, 
\ee
implying that the phase transition is of the third order.

\subsubsection{Another deformation} 
\label{sec:phi4_another}
We consider the same model but with the twist $\alpha$ changed to a $\phi$-dependent function. 
As the function $V(\epsilon\phi)$ in (\ref{deformation}), 
we choose the simplest $Z_2$-symmetric one $V(\epsilon\phi) = (\epsilon\phi)^2$ 
for a technical reason. 

After the limit $\alpha\to 0$, the corresponding saddle point equation (\ref{eq_deform}) is 
\be
\int dy\,\rho(y) {\rm P}\frac{1}{x-y} = x^3+\left(\mu^2-\frac{1}{c_2}\right) x 
\label{eq_phi4_deform} 
\ee
with $c_2 \equiv \int dy\, y^2\rho(y)$. 

\paragraph{One-cut solution} 
To solve the equation (\ref{eq_phi4_deform}), 
we assume the support of $\rho(x)$ as $x\in[-a,a]$ from the 
$Z_2$ symmetry. 
The holomorphic function $F(z) \equiv \int^a_{-a} dy \, \frac{\rho(y)}{z-y}$ is determined as 
\be
F(z) = z^3 + \left(\mu^2-\frac{1}{c_2}\right)z -\left(z^2 +\frac{8-a^4}{4a^2}\right)\sqrt{z^2-a^2}
\ee
with
\be
c_2 = \frac{a^2}{32}(a^4+8).
\ee
Then, 
\be
\rho(x) = \frac{1}{\pi}\left(x^2 + \frac{8-a^4}{4a^2}\right)\sqrt{a^2-x^2}, 
\ee
and $a$ is given as a solution of the equation 
\be
3a^8 +4\mu^2 a^6+16a^4 +32\mu^2 a^2-192=0. 
\ee
Although the explicit form of $a$ is somewhat complicated, 
it can be seen that $0<a^2\le 2\sqrt{2}$ when $\mu^2\ge -1/\sqrt{2}$. 
Then, $\rho(x)\ge 0$ is satisfied, and the solution is valid.   
SUSY breaking is observed from the expectation value 
\be
\lim_{\alpha\to 0}\left(\lim_{N\to \infty}\vev{\frac{1}{N}\tr \,B}_\alpha\right) 
= -i\left(c_2 +\mu^2\right) \neq 0. 
\ee

In particular, when $\mu=0$, 
we obtain a simple expression of the solution as 
\bea
 & & a^4 = \frac83\left(\sqrt{10}-1\right), 
\qquad c_2 = \frac{1}{3\sqrt{6}}\sqrt{\sqrt{10}-1}\left(\sqrt{10}+2\right), \nn \\
 & & \lim_{\alpha\to 0}\left(\lim_{N\to \infty}\vev{\frac{1}{N}\tr B}_\alpha\right)  = -ic_2 \neq 0. 
\eea

As a relevant quantity to the planar free energy, we compute (\ref{phi4_MM_F_0_deform}): 
\be
F_0 = -\ln c_2 -\int dx \,dy\, \rho(x) \rho(y)\, {\rm P}\ln |x-y| 
+ \int dx \,\frac12 \left(x^2+\mu^2\right)^2 \rho(x). 
\ee
For the one-cut solution, we obtain
\bea
\left.F_0\right|_{\mu^2=-1/\sqrt{2}} & = & -\frac14\ln 2 +\frac58, \nn \\
\left.\frac{d}{d(\mu^2)}F_0\right|_{\mu^2=-1/\sqrt{2}} & = & \frac{1}{\sqrt{2}}, \nn \\
\left.\frac{d^2}{d(\mu^2)^2}F_0\right|_{\mu^2=-1/\sqrt{2}} & = & \frac13, \nn \\
\left.\frac{d^3}{d(\mu^2)^3}F_0\right|_{\mu^2=-1/\sqrt{2}} & = & \frac{8\sqrt{2}}{27}.
\label{f0_onecut}
\eea


\paragraph{Two-cut solution}
For $\mu^2<-1/\sqrt{2}$, the relevant large-$N$ solution has the $Z_2$-symmetric two cuts 
$\Omega \equiv [-b,-a]\cup [a, b]$ ($0<a<b$). 
The solution is 
\bea
 & & c_2 = \frac12\left(-\mu^2+\sqrt{\mu^4+4}\right), \qquad 
a^2 = c_2 -\sqrt{2}, \qquad b^2 = c_2 +\sqrt{2}, \\
 & & \rho(x) = \frac{1}{\pi}|x| \sqrt{(x^2-a^2)(b^2-x^2)}. 
\eea
SUSY breaking is seen from 
\be
\lim_{\alpha\to 0}\left(\lim_{N\to \infty}\vev{\frac{1}{N}\tr \,B}_\alpha\right) 
= -\frac{i}{2} \left(\mu^2+\sqrt{\mu^4+4}\right)\neq 0. 
\ee

Computing the free energy $F_0$ for the two-cut solution, we find that 
$F_0$, $\frac{d}{d(\mu^2)}F_0$ and 
$\frac{d^2}{d(\mu^2)^2}F_0$ at $\mu^2=-1/\sqrt{2}$ coincide with (\ref{f0_onecut}), but that
\be
\left.\frac{d^3}{d(\mu^2)^3}F_0\right|_{\mu^2=-1/\sqrt{2}} = \frac{4\sqrt{2}}{27}
\ee
is different. 
Thus, the phase transition at $\mu^2=-1/\sqrt{2}$ is of the third order.

\paragraph{Remarks}
Comparing the result here with the one obtained for the constant twist, 
we see that the change of the deformation, namely the constant twist $\alpha$ changed 
to the $\phi$-dependent function $\alpha\frac{1}{N}\tr\left((\epsilon\phi)^2\right)$, 
modifies the precise form of the one- and two-cut solutions for $\rho(x)$, but their qualitative behavior 
and the order of the phase transition remain unchanged. 
It is considered that, 
because in the slowly varying limit the external field is dominant in the fermion bilinear terms, 
the model will have some sensitivity to the form of the twist   
even after it is turned off.

\section{Large-$N$ $\phi^6$ SUSY matrix model under slowly varying potential} 
\label{sec:phi6_MM}
\setcounter{equation}{0}
In this section, we consider the SUSY matrix model with the derivative of the 
superpotential
\be
W'(\phi)= h(\epsilon\phi)=(\epsilon x)^3 +\mu^2(\epsilon x).
\ee  
For finite $N$ and finite $\epsilon$, it can be seen that 
SUSY is not broken, 
for example by using the Nicolai mapping. 
We show that even in this system, at the large-$N$ and slowly varying limit ($\epsilon\to 0$), 
the SUSY is spontaneously broken. 

For the constant external field $\alpha$, the corresponding large-$N$ saddle point equation (\ref{SPE}) 
becomes 
\be
\int dy \rho(y)\, {\rm P}\frac{1}{x-y} = \frac12 (x^3+\mu^2 x)(3x^2+\mu^2). 
\ee
The one-cut solution with the support $x\in[-a,a]$ is obtained as 
\be
\rho(x) =  \frac{1}{2\pi}\left(3x^4+A_2x^2+A_0\right)\sqrt{a^2 -x^2}, 
\ee
where 
\be
A_2= 4\mu^2+\frac32 a^2, \qquad A_0 = \mu^4 +2\mu^2 a^2 +\frac98 a^4, 
\ee
and $a$ solves the equation
\be
15a^6 +24\mu^2 a^4 +8\mu^4 a^2 -32 =0.
\ee
Then, $\vev{\frac1N\tr\,B}_\alpha$ vanishes due to the $Z_2$ symmetry, but   
\be
\vev{\frac{1}{N}\tr \,B^2}_\alpha = 1-\int^a_{-a}dx\,(x^3+\mu^2x)^2 \,\rho(x)
\ee
is independent of $\alpha$, and nonzero in general implying spontaneous SUSY breaking. 
This solution is valid for the case $\mu^2 \ge -3^{2/3}$. At the boundary $\mu^2 = -3^{2/3}$, 
the eigenvalue distribution takes the form 
\be
\rho(x) = \frac{3}{2\pi} \left(x^2 - 3^{-1/3}\right)^2 \sqrt{a^2 -x^2} 
\ee
with $a = 2\cdot 3^{-1/6}$. It touches the zero at $x=\pm 3^{-1/6}$, 
and then $1/3$ of the eigenvalues are distributed in each of the three regions 
$[-2\cdot 3^{-1/6}, -3^{-1/6}]$, $[-3^{-1/6}, 3^{-1/6}]$ and $[3^{-1/6}, 2\cdot 3^{-1/6}]$. 
For $\mu^2 < -3^{2/3}$, the solution develops to have three cuts.

\paragraph{$\mu=0$ case}
For the case $\mu=0$, we obtain a simple expression: 
\be
a^2 = \left(\frac{32}{15}\right)^{1/3}, \qquad A_2 = \frac32 a^2, \qquad A_0 = \frac98 a^4, 
\ee
and
\be
\vev{\frac{1}{N}\tr\,B^2}_\alpha = 1-\int^a_{-a}dx\,x^6 \rho(x) = \frac23 \neq 0. 
\ee

\subsection{Another deformation} 
Similarly to section~\ref{sec:phi4_another}, 
we consider the deformation of the twist $\alpha$  
corresponding to $V(\epsilon\phi) = (\epsilon\phi)^2$ in (\ref{deformation}). 

{}From \eqref{eq_deform} the saddle point equation after $\alpha \to 0$ is 
\be
\int dy \, \rho(y)\, {\rm P} \frac{1}{x-y} 
= \frac12 \left[\left(\mu^4 -\frac{2}{c_2}\right)x + 4\mu^2 x^3 +3x^5\right]  
\ee
with $c_2 \equiv \int dy\, y^2\rho(y)$.
Assuming the support of $\rho(x)$ as $x\in [-a, a]$, 
we solve it as 
\be
\rho(x) = \frac{1}{2\pi}\left(3x^4+A_2x^2+A_0\right)\sqrt{a^2-x^2},  
\ee
where 
\bea
A_2 &= & 4\mu^2 +\frac32 a^2, \nn \\
A_0 & = & \frac{4}{a^2}-\frac34 a^4 -\mu^2 a^2, \nn \\
c_2 & = & \frac{1}{256}\left(15 a^8 + 16 \mu^2 a^6 +64 a^2\right). 
\eea
And $a$ is determined by 
\be
\left(15a^6+24\mu^2 a^4+8\mu^4a^2-32\right)\left(15a^6+16\mu^2a^4+64\right) = 16\cdot 256. 
\label{phi6_MM_deform_a}
\ee

\paragraph{$\mu=0$ case}
For $\mu=0$, eq. (\ref{phi6_MM_deform_a}) is easily solved to obtain 
\be
a^2 = \left(\frac{64}{15}\right)^{1/3}, \qquad A_2 = \frac32 a^2, \qquad A_0 = \frac{4}{5a^2}, 
\qquad c_2 = \frac12 a^2. 
\ee
Then, $\vev{\frac1N \tr\,B^n}_\alpha$ with $n$ odd vanish from the $Z_2$ symmetry, but 
\bea
\lim_{\alpha\to 0}\left(\lim_{N\to\infty}\vev{\frac{1}{N}\tr B^2}_\alpha\right) 
& = & 1- \int^a_{-a}dx \,\rho(x) x^6 = 0, \nn \\
\lim_{\alpha\to 0}\left(\lim_{N\to\infty}\vev{\frac{1}{N}\tr B^4}_\alpha\right) 
& = & 2-4\int^a_{-a}dx \,\rho(x) x^6 
    -2\left(\int^a_{-a}dx \,\rho(x) x^3\right)^2+\int^a_{-a}dx \,\rho(x) x^{12} \nn \\
 & = & \frac{34}{225} \neq 0, 
\eea
implying spontaneous SUSY breaking.

\section{Large-$N$ Gaussian SUSY matrix model under slowly varying potential} 
\label{sec:gaussian_MM}
\setcounter{equation}{0}
In the previous sections, we have investigated the SUSY breaking in the large-$N$ and slowly varying 
limit for $W'(\phi)$ being quadratic or cubic. 
Here, we consider the same problem for the Gaussian SUSY matrix model with $T=1$, where $W'(\phi)$ is linear: 
\be
W'(\phi) = h(\epsilon\phi) = \epsilon \phi. 
\ee 

We first consider the action with the external field $\alpha$:  
\be
S_\alpha = N \tr \left[\frac12 B^2 +iB\epsilon\phi 
+\bar{\psi}\left(e^{i\alpha}-1 +\epsilon\right)\psi\right].  
\ee
Because the fermion determinant does not depend on $\phi$ in this case, the partition function 
is essentially same as the one under the PBC ($\alpha=0$): 
\be
Z_\alpha = \left(\frac{e^{i\alpha}-1+\epsilon}{\epsilon}\right)^{N^2} Z_{\alpha=0}.  
\ee
According to (\ref{G_SUSY_B}) in appendix~\ref{app:gaussian}, 
the Gaussian SUSY matrix model preserves the SUSY.   
Thus, the constant twist gives no effect for SUSY breaking.

\subsection{$Z_2$-symmetric deformation}
\label{sec:GMM_Z2}
Now the twist is deformed to (\ref{deformation}) with $V(\epsilon\phi)=(\epsilon\phi)^2$, 
which preserves the $Z_2$ symmetry under the flip of the signs of both $\phi$ and $B$. 

Then, the saddle point equation (\ref{eq_deform}) 
is obtained as 
\be
\int dy\, \rho(y){\rm P}\frac{1}{x-y} = \left(\frac12 -\frac{1}{c_2}\right)x,
\label{saddle_pt_Z2}
\ee 
with $c_2\equiv \int dy\,y^2 \rho(y)$. 
Under the assumption that the support of $\rho(x)$ is $x\in [-a, a]$, the solution is found as 
\be
\rho(x) = \frac{1}{6\pi} \sqrt{a^2-x^2}, \qquad a=2\sqrt{3}, \qquad c_2 = 3.
\ee
Then the SUSY is spontaneously broken, since   
\be
\lim_{\alpha\to 0}\left(\lim_{N\to \infty} \vev{\frac{1}{N}\tr \,B^2}_\alpha\right) = 1-c_2 =-2 \neq 0. 
\ee

\subsection{$Z_2$-breaking deformation}
\label{GMM_Z2B}
We here consider another deformation of the twist to a linear function of $\phi$: 
$\alpha \frac{1}{N}\tr (\epsilon\phi)$, 
corresponding to $V(\epsilon\phi) = \epsilon\phi$, which breaks the $Z_2$ symmetry. 
(We have not considered this kind of deformation for the quadratic or cubic $W'(\phi)$ 
in the previous sections, 
due to the technical complexity.) 
What is interesting in this case is that, 
as we will see, two kinds of spontaneous symmetry breaking 
with respect to the SUSY and the $Z_2$ symmetry take place at the same time. 

The large-$N$ saddle point equation from (\ref{eq_deform0})  
\be
2\int dy\, \rho(y){\rm P}\frac{1}{x-y} + \frac{i\alpha}{1-e^{-i\alpha c_1}} -x=0
\label{saddle_pt0}
\ee
with $c_1 = \int dx \,x \rho(x)$ is reduced to 
\be
\int dy\, \rho(y){\rm P}\frac{1}{x-y} = \frac12 \left( x- \frac{1}{c_1}\right),
\label{saddle_pt}
\ee
after the limit $\alpha\to 0$. 
Assuming the support of $\rho(x)$ as $x\in [a, b]$, it turns out that 
eq.~(\ref{saddle_pt}) has two solutions as
\bea
{\rm I)}: & & a=-1, \quad b=3, \quad c_1=1, \\
{\rm II)}: & & a=-3, \quad b=1, \quad c_1=-1. 
\eea
For both solutions, $\rho(x)$ takes the form 
\be
\rho(x) = \frac{1}{2\pi}\sqrt{(x-a)(b-x)}. 
\ee
The $Z_2$ symmetry is spontaneously broken at each of the saddle points.  
They are connected to each other by the $Z_2$ transformation. 
The semi-circle shape of $\rho(x)$ is not influenced by changing the $Z_2$-symmetric twist 
to the $Z_2$-breaking one.   

\subsubsection{Superselection rule}
In order to see which saddle point should be chosen, 
let us reconsider the saddle point equation~(\ref{saddle_pt0})  
with keeping the linear term with respect to $\alpha$.   

Then, (\ref{saddle_pt0}) becomes 
\be
\int dy \, \rho(y) {\rm P}\frac{1}{x-y} = \frac12 \left(x-\frac{1}{c_1}-2\hat{\alpha}\right)
\label{saddle_pt_hat}
\ee
with $\hat{\alpha}\equiv \frac14 i\alpha$. 
For the purpose of obtaining physically acceptable solutions from this, 
it is legitimate to regard the small deformation parameter $\alpha$ 
as pure imaginary rather than real, and $\hat{\alpha}$ as a real quantity.  
Taking account of $\cO(\hat{\alpha})$ corrections, the above solutions become  
\bea
{\rm I)}: & & a=-1+\hat{\alpha}, \qquad b = 3 + \hat{\alpha}, \qquad c_1= 1+\hat{\alpha}, \label{I}\\
{\rm II)}: & & a = -3 +\hat{\alpha}, \qquad b= 1+ \hat{\alpha}, \qquad c_1= -1 +\hat{\alpha}. \label{II}
\eea

Using  
\bea
-\int_a^b dx \, dy \, \rho(x) \rho(y) \,{\rm P}\ln |x-y| & = & -\int_a^b dx \, \rho(x) \, \ln (x-a) 
-\frac14 \int_a^b dx \, x^2 \,\rho(x) \nn \\
 & & +\left(\frac{1}{2c_1}+\hat{\alpha}\right) (c_1-a) +\frac{1}{4}a^2
\eea
obtained from integrating (\ref{saddle_pt_hat}), the planar free energy (\ref{F_deform}) 
can be calculated as 
\bea
F_\alpha
 &=&  -\ln \left(e^{4\hat{\alpha}c_1}-1\right) \nn \\
 & & +\frac14 +\frac14 \left(\frac{a+b}{2}\right)^2 -\frac{a}{2c_1} +\frac14 a^2 +\hat{\alpha}(c_1-a).
\label{planar_F} 
\eea
Recall that the first term in the right-hand side of (\ref{planar_F}) comes from the fermion determinant 
$\left(e^{4\hat{\alpha}c_1}-1\right)^{N^2}$.  
So, when $N$ is even, the first term should be replaced with $-\ln \left|e^{4\hat{\alpha}c_1}-1\right|$. 

Let us consider the case $N$ even\footnote{For the case $N$ odd, 
$Z'_\alpha$ becomes negative at I) in \eqref{I} when $\hat{\alpha}<0$ 
and at II) in \eqref{II} when $\hat{\alpha}>0$. Because $Z'_\alpha$ is analogous to the Witten index rather than 
the thermal partition function, 
the negative sign of $Z'_\alpha$ would not immediately mean instability of the system.   
If we consider the free energy associated with the absolute value of $Z'_\alpha$ and choose the superselection 
sector by the value of $-\frac{1}{N^2}\ln \left|Z'_\alpha\right|$, 
the conclusion is same as in the case of even $N$ discussed below.}. The expression for the free energy becomes 
\bea
{\rm I)}: & & F_\alpha
=  -\ln |4\hat{\alpha}| +\frac54 -2\hat{\alpha} +\cO(\hat{\alpha}^2) , \\
{\rm II)}: & & F_\alpha
= -\ln |4\hat{\alpha}| +\frac54 +2\hat{\alpha} +\cO(\hat{\alpha}^2) . 
\eea
The values of the free energy coincide up to the order $\cO(\hat{\alpha}^0)$.  
Comparing the $\cO(\hat{\alpha})$ terms, 
we can see that I) is favored when $\hat{\alpha}>0$, while II) is favored when $\hat{\alpha}<0$. 

For the expectation value of $B$, 
\be
\lim_{\alpha\to 0}\left(\lim_{N\to \infty} \vev{\frac{1}{N}\tr B}_\alpha \right)
= -ic_1 =\left\{\begin{array}{cl} -i & \mbox{for  I)} \\ 
                           +i & \mbox{for  II)} , \end{array} \right.                       
\ee
also 
\be
\lim_{\alpha\to 0}\left(\lim_{N\to \infty} \vev{\frac{1}{N}\tr B^2}_\alpha \right) 
= -1 \qquad \mbox{for I) and II)}, 
\ee
implying that the SUSY is broken at either of I) or II). 
In this case, $B$ plays the roles of two kinds of order parameters: 
one is for the SUSY breaking and the other is for the $Z_2$ symmetry breaking.

\subsubsection{SUSY Ward-Takahashi identity}
The SUSY breaking discussed above can also be seen as a break down of the SUSY Ward-Takahashi identity. 
For example, from the $Q$-transformation $Q(\bar{\psi}\phi) = -iB\phi-\bar{\psi}\psi$, 
if the SUSY is not broken, 
the identity 
\be
-i\vev{\frac{1}{N}\tr (B\phi)} = \vev{\frac{1}{N}\tr (\bar{\psi}\psi)}
\label{SUSY_WI}
\ee
is expected to hold. 

In the presence of the external field, we compute the left- and right-hand sides: 
\bea
(\mbox{l.h.s.}) & = & -i\vev{\frac{1}{N}\tr (B\phi)}_\alpha \nn \\
  & = & \frac{1}{Z_\alpha}\int d^{N^2}\phi \, \left(-\frac{\epsilon}{N}\tr \phi^2\right) 
\left(e^{i\alpha\frac{\epsilon}{N}\tr\phi}-1+\epsilon\right)^{N^2}
e^{-N\tr\left(\frac{\epsilon^2}{2}\phi^2\right)}  , \\
(\mbox{r.h.s.}) & = & \vev{\frac{1}{N}\tr (\bar{\psi}\psi)}_\alpha \nn \\
& = & \frac{1}{Z_\alpha} (-1)\int d^{N^2}\phi\, 
\left(e^{i\alpha\frac{\epsilon}{N}\tr\phi}-1+\epsilon\right)^{N^2-1}
e^{-N\tr\left(\frac{\epsilon^2}{2}\phi^2\right)} .     
\eea
When $N$ is finite, $Z_\alpha \to 1$ as $\alpha\to 0$. (See (\ref{ZG_1}) with $g_1=\epsilon>0$.) 
Thus, the identity holds as $(\mbox{l.h.s.})=(\mbox{r.h.s.}) =-1/\epsilon$. 

On the other hand, taking the large-$N$ limit with $\epsilon \to 0$, and then $\alpha\to 0$ finally, 
we have 
\bea
\epsilon\times (\mbox{l.h.s.}) & \to & -\int dx \, x^2\rho(x) = -2, \\
\epsilon\times (\mbox{r.h.s.}) & = & - \frac{\epsilon}{e^{i\alpha c_1}-1} \to 0, 
\eea 
implying the SUSY breaking. $\epsilon\times (\mbox{l.h.s.})$ jumps from $-1$ to $-2$ at $N=\infty$, 
while $\epsilon\times (\mbox{r.h.s.})$ jumps oppositely from $-1$ to $0$ at $N=\infty$.

\subsection{General deformation}
\label{GMM_general}
In order to see the response of the model under further deformations, 
in this subsection we consider a general deformation of the model. 
The large-$N$ saddle point equation (\ref{eq_deform}) takes the form 
\be
\int dy \, \rho(y) \,{\rm P}\frac{1}{x-y} = \frac12 \left(x - \frac{1}{c}\,V'(x)\right) 
\label{saddle_pt_V}
\ee
for $x$ in the support of $\rho(x)$, with
\be
c\equiv \int dx \,V(x) \,\rho(x). 
\ee

For two simple examples of the $V(x)$: (one is general linear function and the other is cubic), 
large-$N$ solutions are presented.

\subsubsection{Example 1: general linear $V(x)$}
Let us consider the case 
\be
V(x) = v + x  
\label{general_linear}
\ee  
with $v$ constant. Because (\ref{saddle_pt_V}) is invariant under the overall rescaling of $V(x)$, 
any general linear $V(x)$ can be reduced to (\ref{general_linear}).  

Assuming the support of $\rho(x)$ as $x\in [a, b]$, the two solutions are obtained:
\bea
{\rm I)}: & & a=-2+\frac12 (-v+\sqrt{v^2+4}), \qquad b=2+\frac12 (-v+\sqrt{v^2+4}), \nn \\
 & & c=\frac12 (v+\sqrt{v^2+4}), \\
{\rm II)}: & &  a=-2-\frac12 (v+\sqrt{v^2+4}), \qquad b=2-\frac12 (v+\sqrt{v^2+4}), \nn \\
 & &  c=\frac12 (v-\sqrt{v^2+4}), 
\eea 
where 
\be
\rho(x) = \frac{1}{2\pi}\sqrt{(x-a)(b-x)}
\ee
for both cases. These solutions reflect the spontaneous breaking of the $Z_2$ symmetry again. 
For the expectation values of $B$ at the saddle point I) or II), 
\bea
\lim_{\alpha\to 0}\left(\lim_{N\to \infty} \vev{\frac{1}{N}\tr \,B}_\alpha \right) & = & 
-i(c-v) = -i\frac12 \left(-v\pm\sqrt{v^2+4}\right) \neq 0, \\
\lim_{\alpha\to 0}\left(\lim_{N\to \infty} \vev{\frac{1}{N}\tr \,B^2}_\alpha \right) & = & 
-\left(\frac{\mp v+\sqrt{v^2+4}}{2}\right)^2,
\eea
implies SUSY breaking. 

Since $\rho(x)$ and the expectation values of $B$ depend on $v$ even after $\alpha\to 0$, 
we see that, in the large-$N$ and 
slowly varying limit, 
the effect of the constant term $v$ remains even after the external field turned off.  

In the sense of the critical behavior in 
the one-matrix model~\cite{Kazakov:1989bc,Brezin:1990rb,Douglas:1989ve,Gross:1989aw}, 
from the semi-circle shape of $\rho(x)$, 
the solutions represents the $(2,1)$-phase describing the two-dimensional topological gravity.

\subsubsection{Example 2: cubic $V(x)$}
Here, we consider 
\be
V(x) = v + \frac13 x^3
\ee
with $v$ constant. 

Assuming the support of $\rho(x)$ as $x\in [a, b]$, the solution is given as 
\be
a = \sigma -\sqrt{2\sigma (c-\sigma)}, \qquad b = \sigma + \sqrt{2\sigma (c-\sigma)},
\ee
where $\sigma \equiv \frac{a+b}{2}$ and $c$ are determined by 
\bea
 & & \sigma (c-\sigma )(c-2\sigma ) =2c,
\label{c_sigma_eq1} \\
 & & c - v = \frac{\sigma^2}{12}(9c-5\sigma ) -\frac{\sigma^3}{24c}(5c-2\sigma )(c-\sigma)^2. 
\label{c_sigma_eq2}
\eea
The eigenvalue distribution has the form 
\be
\rho(x) = \frac{1}{2\pi c}(c-\sigma -x)\sqrt{(x-a)(b-x)},  
\ee
from which the model is found to belong to the topological gravity phase ($(2,1)-$phase) 
without fine tuning. We will show below that the model can reach the two-dimensional pure gravity phase 
($(2,3)$-phase) by tuning $v$. 
SUSY breaking is indicated by the expectation values of $B$:  
\bea
\lim_{\alpha\to 0}\left(\lim_{N\to \infty} \vev{\frac{1}{N}\tr \,B}_\alpha \right) & = & 
-i\left[\sigma -\frac{\sigma^2}{4c} (c-\sigma)^2\right], \\
\lim_{\alpha\to 0}\left(\lim_{N\to \infty}  \vev{\frac{1}{N}\tr \,B^2}_\alpha \right) & = & 
1-c\sigma+\frac{\sigma^2}{4}(c-\sigma)^2. 
\eea

If we denote the left-hand side of (\ref{c_sigma_eq1}) as $f(\sigma)$, the extrema of $f(\sigma)$ are 
\be
\sigma_*= \frac{3\pm\sqrt{3}}{6}\,c.
\ee 
The $(2,3)$-critical point corresponds to the case that (\ref{c_sigma_eq1}) and (\ref{c_sigma_eq2}) are 
satisfied at $\sigma = \sigma_*$~\cite{Brezin:1977sv}. The following two critical solutions are found. \\ 
\noindent
I) $v_* = \frac23 \cdot 3^{3/4} (13-6\sqrt{3})$ with
\bea
 & & c_* = 2\cdot 3^{3/4}, \qquad \sigma_* = (\sqrt{3}-1)\cdot 3^{1/4}, \nn \\
 & & a_* = -(3-\sqrt{3})\cdot 3^{1/4}, \qquad b_* = (\sqrt{3}+1)\cdot 3^{1/4}, \nn \\
 & & \rho_*(x) = \frac{1}{2\pi}\frac{1}{2\cdot 3^{3/4}} (b_*-x)^{3/2}\sqrt{x-a_*},  
\eea
II) $v_*=-\frac23 \cdot 3^{3/4} (13-6\sqrt{3})$ with
\bea
 & & c_*=-2\cdot 3^{3/4}, \qquad \sigma_* = -(\sqrt{3}-1)\cdot 3^{1/4}, \nn \\
 & & a_* = -(\sqrt{3}+1)\cdot 3^{1/4}, \qquad b_* = (3-\sqrt{3})\cdot 3^{1/4}, \nn \\
 & & \rho_*(x) = \frac{1}{2\pi}\frac{1}{2\cdot 3^{3/4}} (x-a_*)^{3/2}\sqrt{b_*-x}. 
\eea
I) and II) are related by the $Z_2$ transformation. 
The fractional power $3/2$ in the fall-off of $\rho_*(x)$ at the edge $a_*$ or $b_*$ is 
characteristic of the $(2,3)$-critical point~\cite{Kazakov:1989bc,Brezin:1990rb,Douglas:1989ve,Gross:1989aw}. 
The critical behavior describing the two-dimensional 
pure gravity is obtained by approaching $b_*$ at I) ($a_*$ at II)). 

At the critical point, the expectation values of $B$ become
\bea  
 \left.\lim_{\alpha\to 0}\left(\lim_{N\to \infty} \vev{\frac{1}{N}\tr \,B}_\alpha\right) \right|_* & = & 
\mp i\frac{2-\sqrt{3}}{2}\cdot 3^{3/4} 
\qquad \mbox{for }\left\{\begin{array}{c} {\rm I)} \\ {\rm II)}. \end{array}\right. \\
 \left.\lim_{\alpha\to 0}\left(\lim_{N\to \infty} \vev{\frac{1}{N}\tr \,B^2}_\alpha\right) \right|_* & = & 
10 -6\sqrt{3} \qquad \mbox{for I) and II)}. 
\eea  

Similarly, higher critical points of the one-matrix models, 
$(2,2k-1)$-critical points $(k\ge 3)$~\cite{Kazakov:1989bc,Brezin:1990rb,Douglas:1989ve,Gross:1989aw}, 
could be accessed by the deformation with some polynomial $V(x)$ of the degree $2k-1$. 
Our observation that tuning $V(x)$ yields various phases of two-dimensional gravity manifests the fact 
that choosing the twist itself is a part of the definition of the model as stated in introduction.

\setcounter{equation}{0}
\section{Summary and discussion}
\label{sec:summary}
In this paper, we first discussed spontaneous SUSY breaking in the (discretized) 
quantum mechanics. The twist $\alpha$, playing a role of the external field, 
is introduced to detect the SUSY breaking, as well as 
to regularize the supersymmetric partition function (analogous to the Witten index) 
which becomes zero when the SUSY is broken. 
Differently from spontaneous breaking of ordinary (bosonic) symmetry, 
SUSY breaking does not require cooperative phenomena and can take place even in the discretized 
quantum mechanics with a finite number of discretized time steps. 
There is such a possibility, when the supersymmetric partition function vanishes.   
In general, some non-analytic behavior is necessary for spontaneous symmetry breaking 
to take place. 
For SUSY breaking in the finite system, 
it can be understood that the non-analyticity comes from the vanishing partition function. 
        
We next gave an overview of a class of large-$N$ SUSY matrix models with the slowly varying 
parameter $\epsilon$ introduced. 
The twist $\alpha$ is also introduced to detect SUSY breaking, and it is turned off at the final step. 
In the large-$N$ limit with $\epsilon$ kept finite, there is a possibility that the SUSY, 
which is broken for finite $N$, becomes restored at large $N$. 
On the other hand, it was found that, in the large-$N$ limit and the slowly varying limit $\epsilon\to 0$, 
the matrix models undergo spontaneous SUSY breaking for general superpotential.   
In the slowly varying limit, the fermion interaction terms become negligible compared with 
the term of the external field, which is the origin of the SUSY breaking. 
Furthermore, the superselection rule holds by taking the large-$N$ limit. 
In the finite systems, because the superselection rule does not work, 
we saw that the expectation values of some variables diverges as $\alpha\to 0$. 
However, in the matrix models, there is no such singular behavior thanks to the large-$N$ limit.  
        
As a concrete example of the restoration of SUSY at large $N$, 
we discussed a SUSY matrix model with a double-well potential (the derivative of the 
superpotential $W'(\phi)$ is quadratic). 
In analogy with the infinite volume limit of quantum field theory, 
SUSY is expected to be broken by the instanton effect for finite $N$, 
but it is to be restored due to the suppression of instantons at infinite $N$. 
We obtained the large-$N$ solutions of the matrix model, 
and explicitly saw a number of supersymmetric behavior ensuring the restoration of the SUSY. 

In the large-$N$ and slowly varying limit in the case of $W'(\phi)$ quadratic or cubic, 
we observed spontaneous SUSY breaking, 
and when changing the constant twist   
to a $\phi$-dependent function (proportional to $\tr \,\phi^2$), we also found SUSY breaking. 
The change of the twist does not deform the eigenvalue distribution qualitatively, 
but modifies its precise form. 
Because the external field becomes dominant in the fermion bilinear terms 
in the slowly varying limit, 
the model will remain somewhat sensitive to the form of the twist even after turning it off. 
Thus, we can say that the model is not fully defined by the matrix model action alone, 
and the form of the twist should also be specified to determine the model.  
For the Gaussian SUSY matrix model, where $W'(\phi)$ is linear, the constant twist   
essentially changes nothing from the model under the PBC. 
But, we explicitly saw that some $\phi$-dependent twists cause SUSY breaking. 
 
For future directions, because the twists introduced in the models seem somewhat related   
to the Wilson line of 
an external gauge field, 
the models might be reformulated by introducing some dynamical gauge fields. 
Then, if the external field is dynamically generated 
by the expectation value of the gauge field, 
the models with various twist potentials $V(\epsilon\phi)$ could be unified to a single model 
with gauge interactions.   

It is certainly interesting to discuss whether the mechanism of SUSY breaking 
in this paper is applicable to 
the reduced super Yang-Mill matrix models including the IIB matrix model~\cite{Ishibashi:1996xs}. 
For these models, 
the partition function 
is computed in~\cite{Moore:1998et} to be nonzero for any finite $N$. 
Thus, it is hard to expect that the SUSY is spontaneously broken for finite $N$, 
and in particular for the IIB matrix model 
it would be natural to consider a possibility of SUSY breaking  
at large $N$, if we believe 
that the IIB matrix model describes our real world. 
Since such models have flat directions, for this purpose, it would be an interesting step 
to lift (some of) the flat directions slightly by introducing the slowly varying parameter $\epsilon$ 
with keeping some of the SUSY. 
To carry out the analysis, the method of the Gaussian expansion or improved perturbation theory  
would be useful~\cite{Oda:2000im,Nishimura:2001sx,Nishimura:2002va,Nishimura:2003gz,Aoyama:2006yx}.

\section*{Acknowledgements}
We would like to thank Shinji~Hirano, Hikaru~Kawai, Shoichi~Kawamoto, So~Matsuura, 
Jun~Nishimura, Hidehiko~Shimada, Hiroshi~Suzuki and Asato~Tsuchiya for useful discussions. 
Also, one of the authors (F.~S.) thanks the members of the Niels Bohr Institute 
for hospitality during his stay, when an initial stage of this work was done. 
The authors thank the Yukawa Institute for Theoretical Physics at Kyoto University. 
Discussions during the YITP workshop YITP-W-09-04 
on ``Development of Quantum Field Theory and String Theory'' were useful to complete this work. 
The work of T.~K. is supported in part by Rikkyo University Special Fund for Research, 
and the work of F.~S. is supported in part by a Grant-in-Aid for Scientific Research (C), 21540290.

\appendix
\section{Partition function with the twist $\alpha$}
\label{app:TPF}
\setcounter{equation}{0}
In this appendix, we show that the partition function with the TBC  
for the fermions (\ref{twistedBC}) can be expressed in the form (\ref{partition_fun_TBC}).

Let 
$\hat{b}, \hat{b}^\dagger$ be annihilation and creation operators of the fermions: 
\be
\hat{b}^2 = (\hat{b}^\dagger)^2 = 0, \qquad \{ \hat{b}, \hat{b}^\dagger \} =1, 
\ee
and they are represented on the Fock space $\{ |0\ket, |1\ket \}$ as 
\be
\hat{b}|0\ket =0, \qquad \hat{b}^\dagger |0\ket = |1\ket .
\ee
We assume that $|0\ket, |1\ket$ have the fermion numbers $F=0, 1$, respectively.  

The coherent states $|\psi\ket, \bra \bar{\psi}|$ satisfying
\be
\hat{b} |\psi\ket = \psi |\psi\ket, \qquad 
\bra\bar{\psi}| \hat{b}^\dagger = \bra \bar{\psi}| \bar{\psi}
\ee
($\psi, \bar{\psi}$ are Grassmann numbers, and anticommute with $\hat{b}, \hat{b}^\dagger$.) 
are explicitly constructed as 
\bea
|\psi\ket & = & |0\ket -\psi |1\ket = e^{-\psi\hat{b}^\dagger}|0\ket, \nn \\
\bra \bar{\psi}| & = & \bra 0| -\bra 1|\bar{\psi} = \bra 0| e^{-\hat{b}\bar{\psi}}. 
\eea
Also, 
\bea
|0\ket = \int d\psi \, \psi|\psi\ket, & & \bra 0| = \int d\bar{\psi} \, \bra \bar{\psi}| \bar{\psi}, \nn \\
|1\ket = -\int d\psi \, |\psi\ket, & & \bra 1| = \int d\bar{\psi} \, \bra \bar{\psi}|. 
\eea

Thus, we can obtain 
\bea
\Tr \left[(-e^{-i\alpha})^F e^{-\beta H}\right] & = & 
\bra 0| e^{-\beta H} |0\ket -e^{-i\alpha}\bra 1| e^{-\beta H}|1\ket \nn \\
 & = & \int d\bar{\psi}d\psi \, (e^{-i\alpha} +\psi\bar{\psi})\bra\bar{\psi}| e^{-\beta H}|\psi\ket \nn \\
 & = & e^{-i\alpha} \int d\bar{\psi}d\psi \, e^{e^{i\alpha}\psi\bar{\psi}} \bra \bar{\psi}| e^{-\beta H} |\psi\ket. 
\label{twistedTr} 
\eea

Since the bosonic part of $H$ is obvious, below we focus on the fermionic part 
$H_F = \hat{b}^\dagger W'' \hat{b}$. 
Dividing the interval $\beta$ into $M$ short segments of length $\varepsilon$: 
$\beta = M\varepsilon$ in (\ref{twistedTr}) and applying the relations 
\be
\bra\bar{\psi}|\psi\ket = e^{\bar{\psi}\psi}, \qquad 1=\int d\bar{\psi}d\psi \, |\psi\ket e^{\psi\bar{\psi}}\bra\bar{\psi}|
\ee
to each segment, we have the following expression: 
\bea
\Tr \left[(-e^{-i\alpha})^F e^{-\beta H_F}\right] & = & 
-e^{-i\alpha} \int \left(\prod_{j=1}^M d\psi_j d\bar{\psi}_j\right) \nn \\
 & & \times \exp\left[-\varepsilon \sum_{j=1}^M \bar{\psi}_j\left(\frac{\psi_{j+1}-\psi_j}{\varepsilon} + W''\psi_j\right)\right]
\eea
with 
\be
\psi_{M+1} = e^{i\alpha}\psi_1,  
\label{twBC1}
\ee
or
\bea
\Tr \left[(-e^{-i\alpha})^F e^{-\beta H_F}\right] & = & -e^{-i\alpha} \int \left(\prod_{j=1}^M d\psi_j d\bar{\psi}_j\right) \nn \\
 & & \times \exp\left[-\varepsilon \sum_{j=1}^M \left(-\frac{\bar{\psi}_j-\bar{\psi}_{j-1}}{\varepsilon} + \bar{\psi}_j W''\right) \psi_j\right]
\eea
with 
\be
\bar{\psi}_0 = e^{i\alpha} \bar{\psi}_M. 
\label{twBC2}
\ee
Since (\ref{twBC1}) and (\ref{twBC2}) correspond to (\ref{twistedBC}) 
in the continuum limit $\varepsilon\to 0, M \to \infty$ with $\beta=M\varepsilon$ fixed, 
we find that the formula (\ref{partition_fun_TBC}) holds.

\section{Gaussian SUSY matrix model}
\label{app:gaussian}
\setcounter{equation}{0}
In this appendix, 
we compute the partition function of the Gaussian SUSY matrix model with $T=1$ under the PBC, 
and see that the SUSY is not spontaneously broken. 
{}From \eqref{S_TBC} and \eqref{Z_TBC} with $\alpha=0$, the action and the partition function are defined by 
\bea
S_G & \equiv & N\tr \left[\frac12 B^2 +ig_1 B\phi +g_1 \bar{\psi}\psi\right], \\
Z_G & \equiv & (-1)^{N^2}\int d^{N^2}B \,d^{N^2}\phi \,d^{N^2}\psi \,d^{N^2}\bar{\psi}\, e^{-S_G}.  
\label{ZG_def}
\eea
The corresponding superpotential is given by $W'(\phi) = g_1\phi$, 
and $g_1$ is a (positive or negative) coupling constant.   

{}Under the normalization (\ref{normalization}), it is easy to see 
\be
Z_G = \left({\rm sgn}(g_1)\right)^{N^2} = \left({\rm sgn}(g_1)\right)^{N} 
\label{ZG_1}
\ee
for any $N$. 
On the other hand, the partition function at large $N$ is evaluated by a saddle point configuration of 
the eigenvalue distribution of $\phi$:
\be
\rho_G(x) = \frac{1}{N} \,\tr \,\delta(x-\phi)
\ee
as 
\bea
Z_G  & = &   \left({\rm sgn}(g_1)\right)^N 
\,C_N \,\exp\left\{N^2\left[\int dx\, dy \rho_G(x)\rho_G(y)\,{\rm P}\ln|x-y| 
-\int dx \,\frac12 x^2\rho_G(x)\right]\right. \nn \\
 & & \left. \hspace{3.7cm} \frac{}{} +\cO(N^0)\right\} .
\label{ZG_2}
\eea
$C_N$ is a numerical factor whose large-$N$ behavior is fixed below. 

Using the standard method in \cite{Brezin:1977sv}, it turns out that $\rho_G(x)$ has a semi-circle shape 
in the support $x\in [-2,2]$:  
\be
\rho_G(x) = \frac{1}{2\pi}\sqrt{4-x^2}.
\ee
The exponential factor of (\ref{ZG_2}) can be explicitly computed to give 
\be
 Z_G =  \left({\rm sgn}(g_1)\right)^N \,C_N \,\exp\left[-\frac34N^2 +\cO(N^0)\right]. 
\ee
Comparing this with (\ref{ZG_1}), the large-$N$ behavior of $C_N$ is determined as 
\be
 C_N = \exp\left[\frac34N^2 +\cO(N^0) \right]. 
\label{C_N} 
\ee

Since the $\phi$-integrals in (\ref{ZG_def}) yield the delta-function of $B$, we can see that 
all the expectation values of $\frac{1}{N}\tr \,B^n$ ($n=1,2,\cdots$) vanish 
\be
\vev{\frac{1}{N}\tr\,B^n} = 0, 
\label{G_SUSY_B}
\ee
which implies that the SUSY is not spontaneously broken. 
 
\section{Proof of (\ref{Bn=0})}
\label{app:proof}
\setcounter{equation}{0}
 In this appendix, we give a proof of (\ref{Bn=0}), which serves as a strong evidence for the restoration 
of SUSY in the large-$N$ matrix model with a double-well potential discussed 
in section~\ref{sec:phi4_one-cut}. 
  
First, (\ref{Bn=0}) holds for $n=1,2$ as seen by the explicit computation. 

Next, let us fix an arbitrary $n\ge 2$.  
In order to prove (\ref{Bn=0}) by induction, 
it is sufficient to show 
$\vev{\frac{1}{N}\tr \,B^{n+1}}=0$ under the assumption 
\be
\vev{\frac1N\tr \,B^k}=0 \qquad (k=1,2,\cdots, n). 
\ee
In this appendix, for notational simplicity, 
we write neither the symbols $\lim_{\alpha \to 0}$, $\lim_{N\to \infty}$ nor the suffix $\alpha$ of  
the expectation values.  

For our purpose it is useful to notice a Schwinger-Dyson equation for $k\ge 1$, 
$m \ge 0$,
\bea
0 & = & \sum_{l=0}^{k-1} \vev{\frac{1}{N} \tr \,B^{k-1-l}} \vev{\frac{1}{N} \tr \left(B^l \,W'(\phi)^m\right)} \nn \\
  & & -\vev{\frac{1}{N} \tr \left(B^{k+1}\,W'(\phi)^m\right)} -i\vev{\frac{1}{N}\tr\left(B^k \,W'(\phi)^{m+1}\right)},
\label{SDeq}  
\eea
derived by taking the large-$N$ limit of the identity
\be
0 = \int d^{N^2}\phi \, d^{N^2}B \, d^{N^2}\psi \, d^{N^2}\bar{\psi} \, \sum_{\beta=1}^{N^2} 
\frac{\partial}{\partial B_\beta} \,\tr \left[B^k t^\beta W'(\phi)\right] e^{-S_{\alpha}},
\ee
where $t^{\beta}$ is an appropriate basis of the $N\times N$ hermitian matrices. 
Then, by the assumption, eq.~(\ref{SDeq}) is reduced to 
\be
\vev{\frac{1}{N}\tr\,B^{n+1}} = -i \vev{\frac{1}{N}\tr\left(B^n \,W'(\phi)\right)}
\label{SDeq2}
\ee
for $k=n$, $m=0$, and 
\be
\vev{\frac{1}{N}\tr\left(B^{k+1} \,W'(\phi)^m\right)} = 
-i \vev{\frac{1}{N}\tr\left(B^{k} \,W'(\phi)^{m+1}\right)} +
\vev{\frac{1}{N}\tr\left(B^{k-1} \,W'(\phi)^m\right)}
\label{SDeq3}
\ee
for $1\le k \le n-1$, $m \ge 1$. 

By using these equations repeatedly, we can lower the power of $B$ in $\vev{\frac1N\tr \,B^{n+1}}$, and 
eventually $\vev{\frac1N\tr \,B^{n+1}}$ is expressed as a linear combination of $\vev{\frac1N\tr \,W'(\phi)^m}$ for $m=0,\cdots , n+1$.  
The coefficients of the linear combination 
are given by the combinatorics of how many times we use the first and the second term 
on the right-hand side in \eqref{SDeq3}. Here, it is convenient to associate a monomial $s^{k+1}t^m$ 
with $\vev{\frac{1}{N}\tr\left(B^{k+1} \,W'(\phi)^m\right)}$. Then, \eqref{SDeq2} and \eqref{SDeq3} 
tell that we can make the following replacement: 
\bea
s^{n+1} & \Rightarrow & s^{n+1}\times(-i)\frac{t}{s} , \nn \\
s^{k+1}t^m & \Rightarrow & s^{k+1}t^m \times \left((-i) \frac{t}{s} + \frac{1}{s^2}\right).  
\eea
Then, $\vev{\frac1N\tr \,B^{n+1}}$ is eventually associated with $s^0$-terms 
by making these replacements repeatedly starting from $s^{n+1}$. 
It amounts to the $s^0$-terms of 
\be
-is^nt \times \sum_{m=0}^\infty \left((-i) \frac{t}{s} + \frac{1}{s^2}\right)^m 
= -is^nt \,\frac{1}{1+i\frac{t}{s} -\frac{1}{s^2}},
\ee 
which can be expressed as  
\be
-i \oint_C \frac{ds}{2\pi i} \,\frac{1}{s} \,\frac{s^nt}{1+i\frac{t}{s} -\frac{1}{s^2}} 
= -i \oint_C \frac{ds}{2\pi i} \frac{s^{n+1}t}{s^2+ist-1},
\ee 
where $C$ is a sufficiently large circle around the origin. It is easy to compute 
this integral as 
\be
\frac{-it}{\sqrt{4-t^2}}\left[\left(\frac{-it+\sqrt{4-t^2}}{2}\right)^{n+1} 
-\left(\frac{-it-\sqrt{4-t^2}}{2}\right)^{n+1}\right]. 
\ee
(It turns out to be a polynomial of $t$ of the degree $n+1$.)

Notice that $t$ corresponds to $W'(\phi)$. Using 
\bea
\vev{\frac{1}{N}\tr \,W'(\phi)^m} & = &  \int^b_a dx \,\rho(x) (x^2+\mu^2)^m \nn \\
 & = & \int^2_{-2}\frac{dt}{2\pi}\,\sqrt{4-t^2} \,t^m, 
\eea
we find 
\be
\vev{\frac{1}{N}\tr\,B^{n+1}} =  -i \int^2_{-2} \frac{dt}{2\pi}\,t 
\left[\left(\frac{-it+\sqrt{4-t^2}}{2}\right)^{n+1} 
-\left(\frac{-it-\sqrt{4-t^2}}{2}\right)^{n+1}\right].
\ee 
It is straightforward to see this integral vanishes for all $n\ge 2$.  
Thus, the induction argument completes to prove (\ref{Bn=0}). 



\begin{thebibliography}{999}
{\small 
\bibitem{Banks:1996vh}
  T.~Banks, W.~Fischler, S.~H.~Shenker and L.~Susskind,
  ``M theory as a matrix model: A conjecture,''
  Phys.\ Rev.\  D {\bf 55} (1997) 5112
  [{\tt arXiv:hep-th/9610043}].

\bibitem{Ishibashi:1996xs}
N.~Ishibashi, H.~Kawai, Y.~Kitazawa and A.~Tsuchiya,
 ``A large-N reduced model as superstring,''
  Nucl.\ Phys.\  B {\bf 498} (1997) 467
  [{\tt arXiv:hep-th/9612115}].
  
\bibitem{Dijkgraaf:1997vv}
  R.~Dijkgraaf, E.~P.~Verlinde and H.~L.~Verlinde,
  ``Matrix string theory,''
  Nucl.\ Phys.\  B {\bf 500} (1997) 43
  [{\tt arXiv:hep-th/9703030}].

\bibitem{Kuroki:2007iy}
  T.~Kuroki and F.~Sugino,
  ``Spontaneous Supersymmetry Breaking by Large-N Matrices,''
  Nucl.\ Phys.\  B {\bf 796} (2008) 471
  [{\tt arXiv:0710.3971 [hep-th]}].
  

\bibitem{Witten:1982df}
E.~Witten,
``Constraints On Supersymmetry Breaking,''
  Nucl.\ Phys.\  B {\bf 202} (1982) 253.
  
\bibitem{Mermin:1966fe}
  N.~D.~Mermin and H.~Wagner,
  ``Absence of ferromagnetism or antiferromagnetism in one-dimensional or 
two-dimensional isotropic Heisenberg models,''
  Phys.\ Rev.\ Lett.\  {\bf 17} (1966) 1133.
  
\bibitem{Coleman:1973ci}
  S.~R.~Coleman,
  ``There are no Goldstone bosons in two-dimensions,''
  Commun.\ Math.\ Phys.\  {\bf 31} (1973) 259.
  
\bibitem{Witten:1981nf}
  E.~Witten,
  ``Dynamical Breaking Of Supersymmetry,''
  Nucl.\ Phys.\  B {\bf 188} (1981) 513.

  
\bibitem{Affleck:1983pn}
I.~k.~Affleck,
``Supersymmetry Breaking At Large N,''
  Phys.\ Lett.\  B {\bf 121} (1983) 245.




\bibitem{kanamori-ss} 
  I.~Kanamori, H.~Suzuki and F.~Sugino,
  ``Euclidean lattice simulation for the dynamical supersymmetry breaking,''
  Phys.\ Rev.\  D {\bf 77} (2008) 091502
  [{\tt arXiv:0711.2099 [hep-lat]}];
%
  ``Observing dynamical supersymmetry breaking with euclidean lattice simulations,''
  Prog.\ Theor.\ Phys.\  {\bf 119} (2008) 797 
[{\tt arXiv:0711.2132 [hep-lat]}].
  
\bibitem{catterall4}
S.~Catterall,
``Lattice supersymmetry and topological field theory,''
JHEP {\bf 0305} (2003) 038
[{\tt arXiv:hep-lat/0301028}].


\bibitem{Nicolai:1979nr}
  H.~Nicolai,
  ``On A New Characterization Of Scalar Supersymmetric Theories,''
  Phys.\ Lett.\  B {\bf 89} (1980) 341.




\bibitem{Brezin:1990rb}
  E.~Brezin and V.~A.~Kazakov,
  ``EXACTLY SOLVABLE FIELD THEORIES OF CLOSED STRINGS,''
  Phys.\ Lett.\  B {\bf 236} (1990) 144.

\bibitem{Douglas:1989ve}
  M.~R.~Douglas and S.~H.~Shenker,
  ``Strings In Less Than One-Dimension,''
  Nucl.\ Phys.\  B {\bf 335} (1990) 635.

\bibitem{Gross:1989aw}
  D.~J.~Gross and A.~A.~Migdal,
  ``Nonperturbative Two-Dimensional Quantum Gravity,''
  Phys.\ Rev.\ Lett.\  {\bf 64} (1990) 127;
  ``A NONPERTURBATIVE TREATMENT OF TWO-DIMENSIONAL QUANTUM GRAVITY,''
  Nucl.\ Phys.\  B {\bf 340} (1990) 333.
  

  
\bibitem{Kostov:1988fy}
  I.~K.~Kostov,
  ``$O(n)$ vector model on a planar random lattice: spectrum of anomalous dimensions,''
  Mod.\ Phys.\ Lett.\  A {\bf 4} (1989) 217.
  
\bibitem{Kostov:1992pn}
  I.~K.~Kostov and M.~Staudacher,
  ``Multicritical phases of the $O(n)$ model on a random lattice,''
  Nucl.\ Phys.\  B {\bf 384} (1992) 459
  [{\tt arXiv:hep-th/9203030}].

\bibitem{Eynard:1995nv}
  B.~Eynard and C.~Kristjansen,
  ``Exact Solution of the $O(n)$ Model on a Random Lattice,''
  Nucl.\ Phys.\  B {\bf 455} (1995) 577
  [{\tt arXiv:hep-th/9506193}].

\bibitem{Brezin:1977sv}
  E.~Brezin, C.~Itzykson, G.~Parisi and J.~B.~Zuber,
  ``Planar Diagrams,''
  Commun.\ Math.\ Phys.\  {\bf 59} (1978) 35.

  


\bibitem{Cicuta:1986pu}
  G.~M.~Cicuta, L.~Molinari and E.~Montaldi,
  ``Large N Phase Transitions In Low Dimensions,''
  Mod.\ Phys.\ Lett.\  A {\bf 1} (1986) 125.

\bibitem{Nishimura:2003gz}
  J.~Nishimura, T.~Okubo and F.~Sugino,
  ``Testing the Gaussian expansion method in exactly solvable matrix models,''
  JHEP {\bf 0310} (2003) 057
  [{\tt arXiv:hep-th/0309262}].


\bibitem{Kazakov:1989bc}
  V.~A.~Kazakov,
  ``The Appearance of Matter Fields from Quantum Fluctuations of 2D Gravity,''
  Mod.\ Phys.\ Lett.\  A {\bf 4} (1989) 2125.


\bibitem{Moore:1998et}
  G.~W.~Moore, N.~Nekrasov and S.~Shatashvili,
  ``D-particle bound states and generalized instantons,''
  Commun.\ Math.\ Phys.\  {\bf 209} (2000) 77
  [{\tt arXiv:hep-th/9803265}].

\bibitem{Oda:2000im}
  S.~Oda and F.~Sugino,
  ``Gaussian and mean field approximations for reduced Yang-Mills integrals,''
  JHEP {\bf 0103} (2001) 026
  [{\tt arXiv:hep-th/0011175}].\\
  F.~Sugino,
  ``Gaussian and mean field approximations for reduced 4D supersymmetric Yang-Mills integral,''
  JHEP {\bf 0107} (2001) 014
  [{\tt arXiv:hep-th/0105284}].

\bibitem{Nishimura:2001sx}
  J.~Nishimura and F.~Sugino,
  ``Dynamical generation of four-dimensional space-time in the IIB matrix model,''
  JHEP {\bf 0205} (2002) 001
  [{\tt arXiv:hep-th/0111102}].\\
  H.~Kawai, S.~Kawamoto, T.~Kuroki, T.~Matsuo and S.~Shinohara,
 ``Mean field approximation of IIB matrix model and emergence of four dimensional space-time,''
  Nucl.\ Phys.\  B {\bf 647} (2002) 153
  [{\tt arXiv:hep-th/0204240}]. \\
H.~Kawai, S.~Kawamoto, T.~Kuroki and S.~Shinohara,
 ``Improved perturbation theory and four-dimensional space-time in IIB matrix model,''
  Prog.\ Theor.\ Phys.\  {\bf 109} (2003) 115
  [{\tt arXiv:hep-th/0211272}]. \\
T.~Aoyama, H.~Kawai and Y.~Shibusa,
``Stability of 4-dimensional space-time from IIB matrix model via improved mean field approximation,''
  Prog.\ Theor.\ Phys.\  {\bf 115} (2006) 1179
  [{\tt arXiv:hep-th/0602244}]. \\
T.~Aoyama and H.~Kawai,
``Higher order terms of improved mean field approximation for IIB matrix model and emergence of four-dimensional space-time,''
  Prog.\ Theor.\ Phys.\  {\bf 116} (2006) 405
  [{\tt arXiv:hep-th/0603146}]. \\
T.~Aoyama and Y.~Shibusa,
 ``Improved perturbation method and its application to the IIB matrix model,''
  Nucl.\ Phys.\  B {\bf 754} (2006) 48
  [{\tt arXiv:hep-th/0604211}].

\bibitem{Nishimura:2002va}
  J.~Nishimura, T.~Okubo and F.~Sugino,
  ``Convergent Gaussian expansion method: Demonstration in reduced Yang-Mills integrals,''
  JHEP {\bf 0210} (2002) 043
  [{\tt arXiv:hep-th/0205253}].

\bibitem{Aoyama:2006yx}
  T.~Aoyama, T.~Kuroki and Y.~Shibusa,
  ``Dynamical generation of non-Abelian gauge group via the improved perturbation theory,''
  Phys.\ Rev.\  D {\bf 74} (2006) 106004
  [{\tt arXiv:hep-th/0608031}].




      
}

\end{thebibliography}
\end{document}